\documentclass[a4paper,11pt]{article}

\usepackage{jheppub}

\usepackage{tikz}
\usetikzlibrary{arrows,shapes,decorations}
\tikzstyle{block}=[draw opacity=0.7,line width=1.4cm]
\tikzstyle{ell} = [draw,fill=NavyBlue!10,thick,ellipse]
\usetikzlibrary{calc}
\usepackage{calc}

\usepackage{fixltx2e}

\usepackage{amssymb,amsmath,amsfonts,amsthm,epsfig,pstricks,pst-coil,graphics,pst-node,pst-plot,psfrag}

\newcommand{\be}{\begin{equation}}
\newcommand{\ed}{\end{equation}}
\def\D{{\cal D}}
\def\I{{\cal I}}
\def\ket#1{|{#1}\rangle}
\def\ketb#1{|{#1}\rangle \! \rangle}
\def\bra#1{\langle{#1}|}
\def\Z{\mathbb{Z}}
\def\Tr{\text{Tr}}
\def\R{\mathbb{R}}
\def\ii{i\textit{\={\i}}}
\def\jj{j\textit{\={\j}}}
\def\kk{k\bar{k}}

\preprint{LMU-ASC 76/15}

\title{Entanglement and topological interfaces}

\author{E. Brehm,}
\author{I. Brunner,}
\author{D. Jaud}
\author{and C. Schmidt-Colinet}
\affiliation{Arnold Sommerfeld Center, Ludwig-Maximilians-Universit\"at\\Theresienstraße 37, 80333 M\"unchen, Germany}

\emailAdd{E.Brehm@physik.uni-muenchen.de}
\emailAdd{Ilka.Brunner@physik.uni-muenchen.de}
\emailAdd{Daniel.Jaud@physik.uni-muenchen.de}
\emailAdd{schmidt.co@physik.uni-muenchen.de}

\abstract{In this paper we consider entanglement entropies in two-dimensional conformal field theories in the presence of topological interfaces. Tracing over one side of the interface, the leading term of the entropy
remains unchanged. The interface however adds a subleading contribution, which can be interpreted as a relative (Kullback-Leibler) entropy with respect to the situation with no defect inserted. Reinterpreting boundaries as topological interfaces of a chiral half of the full theory, we rederive the left/right entanglement entropy in analogy with the interface case. We discuss WZW models and toroidal bosonic theories as examples.}

\begin{document}
\maketitle
\flushbottom

\section{Introduction}\label{sec:Introduction}

In the study of entanglement entropy we are often interested in universal terms, in particular
if these terms depend only on a limited number of parameters in the quantum theory. For two-dimensional
conformal field theories, formulae for such terms were found in the seminal work of~\cite{Holzhey:1994we} 
and~\cite{calabrese_entanglement_2009}. Concretely, consider a critical system with a 
subsystem of dimensionless length $L$ in some units.
For large $L$, the entanglement entropy of this subsystem has the expansion 
\begin{equation}\label{cylinderEE}
S_L = \frac{c}{3}\log L\,+\,S_{\rm sub}\,.
\end{equation}
Here, $c$ is the central charge of the CFT, and the term $S_{\rm sub}$ comprises subleading terms, including 
in particular constant terms. In the context of statistical mechanics, the subleading terms will depend on 
the information of the full statistical system, not just its RG fixed point. In this sense, the subleading 
terms are non-universal, whereas the leading term depends only on basic data of the underlying CFT.

Besides systems defined on Riemann surfaces without boundaries, one area of investigation is centered around 
the entanglement in systems with boundaries or interfaces. There are several possibilities to specify 
subsystems, leading to different entanglement entropies. One possibility is to single out a spacial 
interval terminating at the boundary. If the subsystem of length $L$ ends on a 
boundary specified by some boundary condition $b$, the expression for the entanglement entropy 
becomes~\cite{calabrese_entanglement_2009}
\begin{equation}\label{cylinderEEwithboundary}
S_L^{(b)} = \frac{c}{6}\log L\,+\,\log g_b\,+\,\frac{S_{\rm sub}}{2}\,.
\end{equation}
Comparing with \eqref{cylinderEE}, the factor $1/2$ in the overall coefficient of the leading term reflects 
the area law. The quantity $g_b$ is the universal ground-state degeneracy \cite{PhysRevLett.67.161} of the 
boundary condition $b$. In string theory, $g_b$ defines the mass of the D-brane \cite{Harvey:1999gq}. The 
important observation \cite{calabrese_entanglement_2009} is that 
$\log g_b$ 
in \eqref{cylinderEEwithboundary} is a universal contribution of the boundary. The other subleading terms 
are non-universal, where $S_{\rm sub}$ denotes the same terms as in the bulk case (\ref{cylinderEE}).

Via the folding trick, this result can also be applied to interfaces
between two CFTs if the interface splits the  system symmetrically. In 
\cite{Azeyanagi:2007qj,Takayanagi:2008zz,Estes:2014hka}, both boundary and interface entanglement entropy in 
this real-space case were investigated by AdS/CFT methods based on the Ryu-Takayanagi area law 
\cite{Ryu:2006bv,Ryu:2006ef}. 

Naturally, it is of interest to generalise the results on defect and boundary entropies further. For the 
case of interfaces, one would like to consider situations not constrained by the requirement of geometric 
reflection symmetry.
For the case of boundaries, one would like to consider subsystems that are not specified 
by the geometry of the system, but by decomposing the Hilbert space into left- and 
right-movers~\cite{PandoZayas:2014wsa,Das:2015oha}. In this paper, we will discuss the entanglement entropy 
through the special class of topological interfaces, and show how the same techniques can be employed to 
determine the left/right entanglement entropy for boundaries.

The  problem of entanglement through interfaces has been approached before in special examples, in particular 
for the case of free bosons in   \cite{sakai_entanglement_2008}. The interface splits the system into 
two parts, and one is interested in the entanglement entropy of the subspace on one side of the interface.
The same  method of computation was used in \cite{Brehm:2015lja} to determine the entanglement across 
defects of the Ising model, giving in particular a CFT computation of results obtained earlier 
in~\cite{peschel_entanglement_2005,peschel_entanglement_2010}. A recent investigation from the AdS/CFT point 
of view of these setups can be found in \cite{gutperle:2015hcv}, see \cite{Bak:2011ga,Bak:2013uaa} for further 
results in this direction.

In these examples, it was observed that the entanglement entropy in the presence of interfaces receives 
two kinds of modifications to the pure bulk expression (\ref{cylinderEE}). The first one is a reduction of the 
coefficient of the universal logarithmic term of the bulk theory. This correction depends on the ``strength'' 
of the interface, which can be defined using its transmissivity or reflectivity~\cite{quella_reflection_2007}.  
The other modification is a universal contribution to 
the constant part of the bulk term. This shift was observed to be independent of the transmissivity, and 
indeed only depends on the topological information contained in the interface operator.

In this paper we will study the constant shifts of the entanglement entropy in the presence of 
interfaces in some more detail and generality. Our discussion is restricted to the case of topological 
interfaces, preserving the full conformal invariance of the bulk theory. Topological interfaces 
are tensionless domain walls of two-dimensional CFTs and can be moved freely on the two-dimensional 
space-time as long as no other interfaces or operator insertions are crossed. They add interesting 
structure to any CFT, as they can for example merge smoothly, defining a product structure on the 
space of interfaces. Topological interfaces define maps between the Hilbert spaces of the CFTs on the two sides 
of the interface. These maps in particular intertwine the action of the Virasoro algebra, as a 
consequence of the fact that the interface is topological.  We will consider the case where the interface 
also intertwines the action of higher symmetry generators.

Of particular interest are cases where the higher symmetry renders the theory rational.
Tracing over one side of the topological interface, we show that one obtains a natural reduced density matrix.
Strictly speaking, the associated probability distribution is determined by the fusion product of the interface 
with its conjugate. This combination can in particular be regarded as a map from a single CFT to itself.  
In the case of a rational theory, the probability distribution acts on the space of irreducible representations 
appearing in the decomposition of the bulk Hilbert space of that theory with respect to the symmetry algebra. 
We compute the entanglement entropy through the interface and show that it is given by the negative of the 
Kullback-Leibler entropy relative to the probability distribution associated to the trivial (invisible) interface.

The Kullback-Leibler divergence of one probability distribution with respect to another is always 
nonnegative, and vanishes only if the two probability distributions agree. We discuss what this
means in terms of the interface. The map between interfaces and probability distributions is not 
one-to-one, there is in general more than one interface corresponding to a single probability 
distribution. In particular, not only the trivial interface leads to a vanishing subleading 
contribution to the entanglement entropy, but so do all interfaces corresponding to symmetries. 
These interfaces map states to symmetry-transformed states, and therefore do not lead to any 
information loss. This minimum of the information loss can only 
be saturated if the partition functions of the  CFTs on the two sides of the interface agree and 
the fusion product of the interface with its dual is the identity. For the other cases, we will 
derive a bound on the subleading term of the entanglement entropy.

We apply our result to the $su(2)_k$ WZW models in the limit of large levels $k$, where the theory 
approaches a sigma-model on $S^3$ with $H$-flux. Here, the probability 
distributions associated to  topological defects become continuous. We compute the shift in 
entanglement entropy for elementary defects in the large-$k$ limit. For generic elementary defects, we 
obtain rational numbers approaching $1$. In certain distinguished cases however the Kullback-Leibler divergence 
develops sharp peaks that can be attributed to divisibility properties involving the level $k$. 

Furthermore, we slightly extend the framework of rational CFTs and discuss also toroidal compactifications 
of free bosons. Here, the situation is particularly simple, as all topological interfaces fall into the 
class of ``duality interfaces'' introduced in \cite{Frohlich:2006ch}. The defining property of this class 
of  interfaces is that fusion with the orientiation reversed interface yields a superposition of symmetry 
defects. Thus, the insertion of this superposition amounts to an orbifold-like projection of the initial 
theory. When one associates a probability, non-invariant states are assigned probability $0$, whereas all 
invariant states are distributed equally. This amounts to a shift in the entanglement entropy by the 
logarithm of the number of symmetry defects appearing in the fusion. For the free boson theory, this 
contribution can be identified as the index of a sublattice of the winding and momentum lattice to 
which the interface couples. This index appears also in the $g$-factor of the interface, such that in 
the special case of bosonic theories the shift in the entanglement entropy agrees with $\log g^2$. 
Note that the shift vanishes for duality defects implementing T-duality transformations, which is as expected 
since there  is no information loss due to T-duality.

Besides the entanglement across topological defects we consider entanglement entropies between 
non-geometrical subsectors for boundaries. As mentioned before, in the presence of a boundary a suggestive 
subsector of the space of states is provided by the chirality 
of the symmetry algebra. Since the boundary gluing conditions couple the left- and the right-moving degrees of 
freedom, the idea is to trace out the holomorphic or antiholomorphic sector of the space of states. The authors 
of~\cite{PandoZayas:2014wsa} studied this left/right entanglement entropy for a free boson CFT. Results for 
general rational CFTs and an interpretation in a gravitational context were obtained in~\cite{Das:2015oha}. 

In this paper, we give an alternative derivation of the results of~\cite{Das:2015oha}. For this, we 
re-interpret the conformal boundary as a defect-like object on the full plane. To do so, we start with a 
theory on the upper half-plane with a boundary condition along the real line. We employ the doubling trick 
to fold the dependence on anti-holomorphic coordinates to the lower half plane. The boundary condition on 
the boundary state can then be interpreted as an intertwining property for an interface operator associated to the 
boundary condition. The computation of the entanglement entropy then resembles the one for the interface 
case. However, the interpretation in terms of a Kullback-Leibler divergence is lost, as there is no natural 
trivial boundary condition that could serve as a reference point. 

Again, we consider $su(2)$ WZW models and toroidal theories as examples. In the special case of tori, 
the left/right entanglement entropy is given in terms of the $g$-factor. 

The plan of the paper is as follows. In section~\ref{sec:TopologicalDefects} we recall some basics
of topological interfaces in two-dimensional CFTs, and fix some notation. In section~\ref{sec:HowTo} 
we briefly review the method of \cite{Holzhey:1994we,sakai_entanglement_2008} for the computation of 
the entanglement entropy through conformal interfaces. We then turn to the case where the interface is 
topological in section~\ref{sec:EErational}. Here, we work out the relation between entanglement entropy and 
Kullback-Leibler divergence, and discuss the rational case in detail. The relevance  of our formulae to 
the left/right entanglement entropy (LREE) is discussed in section~\ref{sec:LREErational}. 
Finally, we turn to the case of free bosons compactified on tori in section \ref{sec:EEtori}. We determine 
the entanglement entropy through topological interfaces as well as the left/right entanglement entropy 
in this case, which slightly extends the rational framework in a simple example.

\section{Topological Interfaces}\label{sec:TopologicalDefects}

The study of one-dimensional interfaces between two-dimensional CFTs has a long 
history~\cite{Oshikawa:1996ww,Bachas:2001vj}. Locally, an interface sets gluing conditions 
for all pairs of local fields separated by the interface. In that sense an interface defines
a map between the algebras of local fields on the two sides. Similar to a boundary
condition, such an interface condition admits local excitations and constitutes a one-dimensional subsector 
of the full quantum field theory.  
At a conformal fixed point the interface preserves at least one half of 
the bulk conformal charges. If the interface runs along the real axis of the complex plane and separates CFT1 from 
CFT2, the condition reads
\begin{equation}\label{eq:interfaceconformality}
\lim_{y\searrow 0}(\,T^{(1)}(x+iy)-\tilde{T}^{(1)}(x-iy)\,)\,=\,\lim_{y\nearrow 0}(\,T^{(2)}(x+iy)-\tilde{T}^{(2)}(x-iy)\,)\,,
\end{equation}
where $T^{(n)}$ and $\tilde{T}^{(n)}$ are the holomorphic and antiholomorphic components of the 
energy-momentum tensor of CFT$n$. The requirement~\eqref{eq:interfaceconformality} is a necessary local 
condition. Similarly as in the case of boundary conditions, further local conditions follow from sewing 
relations, and global conditions arise from modular constraints on the torus. 

In some sense, conformal interfaces generalise the notion of conformal boundary conditions, which are
the special solutions where both sides of~\eqref{eq:interfaceconformality} are equal to zero, or occur 
if one side of the interface is trivial.  Another set 
of special solutions to~\eqref{eq:interfaceconformality} is obtained when the interface commutes with 
both the left- and the right-moving Virasoro algebra, such that~\eqref{eq:interfaceconformality} is 
solved separately for the holomorphic and the antiholomorphic component of the energy-momentum tensor. 
Evidently, this can only happen when the theories on the two sides have equal left- and right-moving
central charges. An interface corresponding to such a solution can be freely deformed 
and moved on the Riemann surface, as long as it does not cross any operator insertions. These interfaces 
were dubbed topological in~\cite{Bachas:2004sy}. Initially introduced in~\cite{Petkova:2000ip}, 
topological interfaces have been studied in particular in rational CFTs (see 
{\it e.g.}~\cite{Frohlich:2006ch,Fuchs:2007tx} and references therein). 

A topological interface can be regarded as an operator on the space of states, acting  as a constant map between (left-right pairs of) 
isomorphic Virasoro representations. In the case where the conformal symmetry is enhanced to a larger 
chiral symmetry algebra, the topological interface condition may or may not respect the additional symmetry. 
Topological interfaces between CFT1 and CFT2 thus naturally fall into classes corresponding to the 
preserved common symmetry subalgebra. For a given topological interface, we consider the 
decomposition of the space of states of CFT$n$ ($n=1,\,2$) with respect to this common subalgebra,
\begin{equation}\label{eq:H-decomposition}
 \mathcal{H}^n = \bigoplus_{(\ii)}M^{n}_{\ii} 
\,\mathcal{H}_i \otimes \mathcal{H}_{\textit{\={\i}}}\,.
\end{equation}
The indices $i$ and $\textit{\={\i}}$ label (generally different) irreducible highest weight representations 
of the two chiral parts of the common subalgebra. The non-negative integers~$M^n_{\ii}$ give the multiplicities 
of the pair of representations~$(i,\textit{\={\i}})$. We will assume that our theories are unitary and have 
a discrete spectrum of highest weight states of the chiral subalgebra, and that there is a unique vacuum state. 
For the following discussion we will also assume that the modular $S$ transformation 
of representation characters is given by a discrete matrix.

An operator corresponding to a general topological interface will then be denoted
\begin{equation}\label{eq:GeneralTopologicalDefect}
\I_A = \sum_{\bf i}d_{A{\bf i}}\,\|{\bf i}\|\,.
\end{equation}
Generally, we will use capital indices to label the interfaces. We use bold-face indices~${\bf i}$
to refer to a pair of left-right products of irreducible representations in the two adjacent CFTs,
\begin{equation}
{\bf i}\equiv (i,\textit{\={\i}}\,;\alpha,\beta)\,.
\end{equation}
Here, $(i,\textit{\={\i}})$ labels the transmitted pair of representations. The indices 
\mbox{$\alpha=1,2,\ldots, M^1_{i\textit{\={\i}}}$} and \mbox{$\beta=1,2,\ldots, M^2_{i\textit{\={\i}}}$} 
are the multiplicity labels of this pair on the two sides of the interface. The symbol~$\|{\bf i}\|$ 
in~\eqref{eq:GeneralTopologicalDefect} denotes the Ishibashi-type projector which acts as an intertwiner 
between the two pairs of representations, {\it i.e.}
\begin{equation}
\,\|{\bf i}\| \, : \mathcal{H}_i \otimes \mathcal{H}_{\textit{\={\i}}}^{(\alpha)}\ 
\to \mathcal{H}_i \otimes \mathcal{H}_{\textit{\={\i}}}^{(\beta)}\
\end{equation}
and
\begin{equation}
J_n \|{\bf i}\| = \|{\bf i}\| J_n \ ,
\end{equation}
where $J_n$ denotes any symmetry generator. Note that in general one can include an automorphism of 
the extended symmetry algebra into the action of the interface; this generalisation is straightforward. 

One important property of topological interfaces is that they admit a fusion product. The fusion 
product has the geometric interpretation of moving the interface lines on top of each other, and 
interpreting the result as a topological interface between the two remaining CFTs. While fusion 
may also be defined for the more general conformal interfaces, it is particularly straightforward 
in the topological case, where it basically consists of map 
composition~\cite{Petkova:2000ip}. When writing the coefficients $d_{AB{\bf i}}$ of the fusion 
product $\I_{AB}=\I_A\I_B$ we will suppress the summation over multiplicity labels,
\begin{equation}
d_{AB{\bf i}}=\sum_{\gamma}d_{A{(i\textit{\={\i}};\alpha,\gamma)}}d_{B{(\ii;\gamma,\beta)}}
\,\equiv\,d_{A{\bf i}}d_{B{\bf i}}\,.
\end{equation}
The fusion product extends to fusion of a topological interface with (non-topological) conformal 
interfaces, in particular with boundary conditions.

Due to invariance under modular $S$ transformation, the coefficients $d_{A{\bf i}}$ must 
satisfy~\cite{Petkova:2000ip}
\begin{equation}\label{eq:integralitycondition}
\sum_{\bf i}S_{ij}S_{\textit{\={\i}}\textit{\={\j}}}\,{\rm Tr}\,d_{A^*{\bf i}}\,d_{A{\bf i}}\,=\,
{\cal N}_{\jj \,A}^{\;\;\;A}\,\in\,\mathbb{N}_0\,,
\end{equation}
where $A^*$ labels the orientation reversed interface with corresponding defect operator $\D_{A^*}=\D_A^\dagger$, $S_{ij}$ is 
an element of the modular $S$ matrix, and the trace is over multiplicity labels. This is the 
analogue of the Cardy condition for conformal boundary states. 
The ${\cal N}_{\jj \,A}^{\;\;\;B}$ count the multiplicity of the pair of representations 
$(j,\textit{\={\j}})$ in a system where the topological interfaces 
intersect a spatial slice, {\it i.e.} for a time evolution parallel to the interfaces. 
The condition \eqref{eq:integralitycondition} restricts the possible values of coefficients 
$d_{A{\bf i}}$, and it also requires that linear superpositions of interfaces 
must have integer coefficients. We refer to interfaces which cannot be decomposed into a 
superposition of other interfaces with positive coefficients as `elementary'. The set of 
elementary interfaces forms a basis for all topological interfaces of the same class. Obviously
any interface for which at least one of the~${\cal N}_{\jj \,A}^{\;\;\;A}$ is equal to 1 is elementary. 
In fact, due to the operator-state correspondence in the theory on the interface any elementary 
interface has at least ${\cal N}_{{00} \,A}^{\;\;\; A}=1$, {\it i.e.} the vacuum in parallel 
time evolution occurs with multiplicity~1.
 
Consider a set of topological interfaces $\I_A$ for which $d_{A{\bf i}}$ provides a unitary 
transformation from projectors $\|{\bf i}\|$ to the $\I_A$. It can be shown \cite{Petkova:2000ip} 
that the corresponding ${\cal N}_{\ii \,A}^{\;\;\;B}$ form a representation of a tensor product 
of fusion algebras,
\begin{equation}\label{eq:NIM-rep}
\sum_{B}{\cal N}_{\ii \, A}^{\;\;\; B}\,{\cal N}_{\jj\, B}^{\;\;\;\,C}\,=
\,\sum_{\bf k}N_{ij}^{\; k}N_{\textit{\={\i}}\textit{\={\j}}}^{\;{\bar{k}}}\,{\cal N}_{\kk\,A}^{\;\;\;\;C}\,.
\end{equation}
In the last formula, the $N_{ij}^{\;k}$ are the fusion rules of the chiral algebra. It is 
easy to see that a topological interface $\I_A$ in such a set is elementary.

A particular instance where we know a set of $d_{A{\bf i}}$ that provides a change of basis occurs
in rational CFT, {\it i.e.} in theories where the index set $\{{\bf i}\}$ in~\eqref{eq:H-decomposition} 
is finite. The simplest case are the diagonal theories --- theories which are charge conjugation 
invariant~($i=\textit{\={\i}}\,$), and where the multiplicities for all chiral algebra representations 
are~1. In such a theory there are topological defects\footnote{In this paper we usually refer to interfaces 
as defects if the CFTs on the two sides are identical.} of the form
\begin{equation}\label{RCFTdefects}
\D_{\bf a}=\sum_{i}\frac{S_{ai}}{S_{0i}}\,\|i\|\,.
\end{equation}
These defects have ${\cal N}_{0a}^{\;a}=N_{0a}^{\;a}=1$ and are therefore elementary. 
They provide a basis for the set of topological defects which respect the chiral symmetry.

In cases where the chiral algebra admits a global symmetry $G$, we find among the topological interfaces 
the so-called symmetry defects. Each element $g\in G$ can be associated to a topological defect $\D_g$. 
By definition, these interfaces glue any field to its image under the symmetry operation.
Hence, they implement an action of $G$ through
\begin{equation}\label{eq:symmetrydefect}
\D_g^\dagger=\D_{g^{-1}}\,,\qquad
\D_g\D_h=\D_{gh}\qquad\forall h,\,g\in G\,. 
\end{equation}
A broader class of interfaces are the duality interfaces introduced in~\cite{Frohlich:2006ch}. 
Their defining property is that
\begin{equation}\label{eq:dualitydefect}
\I\,\I^\dagger=\bigoplus_{g\in G}\D_g\,,
\end{equation}
where $G$ is a finite symmetry group of the CFT. The fusion product of a duality interface with 
its adjoint contains a superposition of group-like defects corresponding to a symmetry (sub-)group. 
Duality defects were first introduced in the context of RCFT, where they can be used to relate CFTs with 
the same chiral algebra but different modular invariants. However, the definition can be extended also 
to the non-rational context. Prominent examples for duality interfaces implement 
dualities such as T-duality in free field theories, or the Kramers-Wannier duality in the Ising 
model~\cite{Frohlich:2004ef}.

\section{Entanglement Entropy}\label{sec:HowTo}

Entanglement entropy measures quantum correlation between subsystems. Let 
$\rho=\vert\psi\rangle\langle\psi\vert$ be the density matrix of a system in a pure quantum state 
$\vert\psi\rangle$. Let the Hilbert space be a direct product $\cal{H} = \cal{H}_\text{A} \otimes 
\cal{H}_\text{B}$ where A and B are the subsystems. The reduced density matrix of A is 
$\rho_{\rm A} = \textnormal{Tr}_{\rm B}\, \rho$. The \textit{entanglement entropy} is the corresponding 
von~Neumann entropy 
\begin{equation}\label{eq:EEdef}
  S_A = -\Tr\, \rho_{\rm A} \log \rho_{\rm A}\,. 
\end{equation}
$S_{\rm B}$ is defined analogously. For the density matrix of a pure quantum state one always has 
$S_{\rm A} = S_{\rm B}$. In the simplest case the pure state is the ground state $\ket{0}$ of the 
system. We then refer to the corresponding quantity~\eqref{eq:EEdef} as the 
\textit{ground state entanglement entropy}. 

One way to compute \eqref{eq:EEdef} makes use of the replica trick \cite{calabrese_entanglement_2009}. The 
theory is considered on $K$ copies of the original Riemann surface, glued together along the subsystem A in a cyclic fashion. 
Tracing over all copies of the subsystem B, the reduced density matrix becomes $\rho_{\rm A}^K$. Its trace in the 
$K$-sheeted Riemann surface can be written as
\begin{equation}
 \text{Tr}_{\rm A}\,\rho_{\rm A}^K = \frac{Z(K)}{Z(1)^K}\,,\label{eq:TrrhoK}
\end{equation}
where $Z(K)$ is the full partition function on the $K$-sheeted Riemann surface.
One analytically continues this expression to complex values of $K$, and obtains the entanglement 
entropy \eqref{eq:EEdef} from
\begin{equation}\label{EEgeneral}
 S_{\rm A} = (1-\partial_K)\log Z(K) \vert_{K\rightarrow1}\,.
\end{equation}

\subsection{The replica trick and conformal interfaces}\label{sec:ReplicaTrickandI}

In the following, we will briefly review a construction of $Z(K)$ due to 
\cite{Holzhey:1994we,sakai_entanglement_2008}, which in principle allows to derive the entanglement entropy 
through general conformal interfaces connecting two conformal field theories. The same construction was 
also used in \cite{Brehm:2015lja}.   

\begin{figure}[h!t]
 \begin{center}
 \begin{tikzpicture} [scale=2]
   \fill [gray!20] (-.4,1.1) rectangle (0,-1.1);
   \draw[->] (-.4,0) -- (1.1,0) node[below left] {\small{Re $w$}};
   \draw[->] (0,-1.1) -- (0,1.1) node[above] {\small{Im $w$}};
   \draw[blue,very thick] (.02,0) -- node[above] {\small{branch cut}} (1,0);
   \draw[very thick,red] (0,-1) -- (0,0);
   \draw[very thick,red] (0,0) -- node[right] {\small{Interface}} (0,1);
   \node at (1.6,0) {$\xrightarrow[\text{cutoffs }\epsilon,\,L]{z = \log w}$};
 \end{tikzpicture}
 \begin{tikzpicture} [scale=2.5]
   \foreach \j in {1,...,5} \fill[gray!20] (-1,2*.\j) rectangle (1,2*.\j-.1);
   \foreach \i in {1,...,9} \draw[red] (-1,.\i)--(1,.\i);
   \draw[->] (-1.1,0) -- (1.1,0) node[right] {\small{Re $z$}};
   \draw[->] (0,-.2) -- (0,1.1) node[above] {\small{Im $z$}};
   \draw[blue] (-1,0) node[below] {\small{$\log \epsilon$}} --  (1,0) 
	node[below] {\small{$\log L$}} -- (1,1) node[right] {\small{$2\pi K$}}-- (-1,1) -- (-1,0);
   \node at (0,-.4) {};
  \end{tikzpicture}
  \vspace{-0.5cm}
 \end{center}
 \caption{\it Sketch of the $K$-sheeted Riemann surface we use in the replica trick. After imposing an 
UV cutoff $\epsilon$, a IR cutoff $L$, the surface corresponds to the cylinder on the upper right. To 
derive $Z(K)$ we impose periodicity also in the direction of the real part of $z$ to obtain a torus.
}
 \label{fig:torus}
\end{figure}
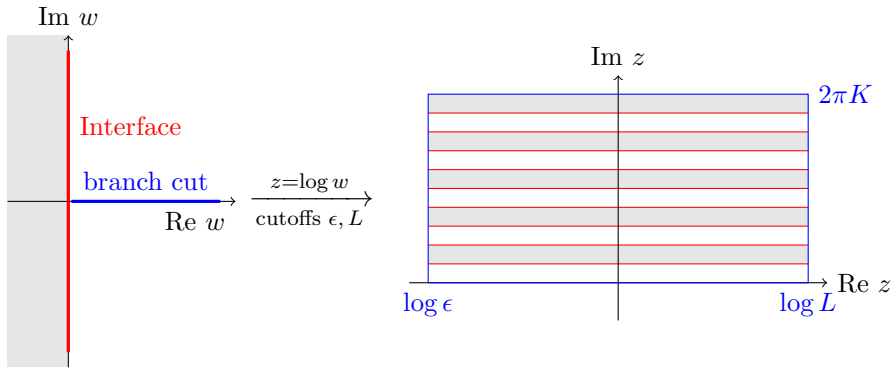

Consider a conformal interface $\I$ along the imaginary axis of the complex plane, with CFT1 on 
$\text{Re}\, w > 0$ and CFT2 on $\text{Re} \,w < 0$. With time flowing along the defect line, the subsystems 
A and B consist of the positive and negative real axis, respectively. Following the replica trick, the 
corresponding $K$-sheeted Riemann surface consists of $K$ copies of the complex plane, glued together cyclically 
along a branch cut on the positive real axis, as illustrated on the left of Figure~\ref{fig:torus}.

In order to evaluate the partition function $Z(K)$ we introduce the cutoffs 
$\vert w \vert =\epsilon$ and $\vert w \vert =L$ and change coordinates to $z=\log w$. Observe that this transformation is
compatible with \eqref{eq:interfaceconformality}. The resulting cylinder is illustrated on the right of 
Figure~\ref{fig:torus}. As in \cite{sakai_entanglement_2008} we regularise the partition function by imposing 
periodicity in ${\rm Re}\,z$ and choosing $\epsilon = \frac{1}{L}$. Periodicity can be imposed since the cut-offs 
$L$ and $\epsilon$ are very large and very small, respectively. In these limits the result does not depend on the 
specific choice of boundary condition. The identification $\epsilon = 1/L$ is somewhat arbitrary. 
It will lead to a factor of two in the final result for the leading term of the entanglement entropy. 
This will have the benefit that the bulk term will have the familiar form $c/3\log L$, even though 
since $L$ is only an IR cut-off, the `interval' is actually physically a half line with only one end point,
in which case the entanglement entropy should be reduced by a factor 1/2 due to the area law.
On the resulting torus one observes that the shape of the defects is unaltered under the global 
conformal transformation that changes the time evolution parallel to the interfaces to the one flowing 
orthogonally to the interfaces. We conclude that $Z(K)$ is given by a torus partition function 
with $2K$ interfaces inserted,
\begin{equation}
\begin{split}
 Z(K) &= \text{Tr}_1 \left(\I^\dagger \,e^{-\delta H_2} \,\I \,e^{-\delta H_1} \cdots 
\I \,e^{-\delta H_1} \right)\\
  &= \text{Tr}_1 \Big(\I^\dagger \,e^{-\delta H_2}\, \I \,e^{-\delta H_1} \Big)^K \,,
\end{split}\label{eq:master1}
\end{equation}
where $H_1$ and $H_2$ are the Hamilton operators in the respective CFT, and 
\begin{equation}\label{def:delta}
\delta = \frac{2 \pi^2}{\log L/\epsilon} = \frac{\pi^2}{\log L}\,.
\end{equation}  
Obviously the evaluation and analytic continuation of \eqref{eq:master1} depends heavily 
on $\I$. For non-topological conformal defects, $Z(K)$ is in general very hard to compute. 
An explicit expression which permitted the computation of the entanglement entropy was 
obtained in \cite{sakai_entanglement_2008} for the case of a single free boson, and 
in~\cite{Brehm:2015lja} for conformal defects of the free fermion and the Ising model. 
However, for topological defects the expression for $Z(K)$ simplifies considerably, as we 
will see in the following.

There is one feature of the entanglement entropy as we define it here which is rather obvious 
already at this stage. From~\eqref{eq:master1} it is easy to see that~\eqref{EEgeneral} is 
invariant under any rescalings of the interface. While interfaces generically have a standard 
normalisation derived from their properties under modular transformations, this means in particular 
that superpositions $M\I$ of identical interfaces $\I$ yield the same entanglement as a single~$\I$.

\section{Entanglement Entropy through Topological Interfaces}\label{sec:EErational}

In the limit of a large IR cutoff $L$, or equivalently $\delta\ll 1$, the entanglement entropy through 
a topological interface $\I$ follows from the torus partition function~\eqref{eq:master1} in a 
rather straightforward way. The torus partition function includes $K$ insertions of $\I$ and of its 
adjoint $\I^\dagger$, which commute with both Virasoro algebras and therefore in particular with 
the Hamiltonian \mbox{$H\propto L_0+\tilde{L}_0$}. Thus we can write
\begin{equation}\label{eq:master2}
 Z(K) = \text{Tr}\left(\I e^{-\delta H}\I^\dagger e^{-\delta H}\right)^K = \text{Tr}\left((\I\I^\dagger)^K 
 e^{-2\delta H K}\right)\,. 
\end{equation}
For the general topological interface \eqref{eq:GeneralTopologicalDefect} we find
\begin{equation}\label{ZforTI}
 Z(K) = \sum_{(i,\textit{\={\i}})} {\rm Tr}\left(d_{A{\bf i}}\,d_{A^*{\bf i}}\right)^K \chi_{i}
\left(e^{-2\delta K}\right)\chi_{\textit{\={\i}}}\left(e^{-2\delta K}\right) \,,
\end{equation}
where $\chi_i(q)$ is the character of the representation $i$. In~\eqref{ZforTI} and in the remainder of this
section, Tr denotes the trace over multiplicity indices. Applying a modular $S$ transformation we obtain
\begin{equation}\label{eq:ZKaftermodularS}
 Z(K) = \sum_{(i,\textit{\={\i}})}\sum_{j,\textit{\={\j}}} {\rm Tr}\left(d_{A{\bf i}}\,d_{A^*{\bf i}}\right)^K 
 S_{ij}S_{\textit{\={\i}}\textit{\={\j}}}
\, \chi_{j}\left(e^{-\frac{2\pi^2}{\delta K}}\right)\chi_{\textit{\={\j}}}\left(e^{-\frac{2\pi^2}{\delta 
K}}\right)\,.
\end{equation}
In the limit $\delta\ll1$ only the vacuum with the energy $E_0=-\frac{c}{12}$ will contribute to the sum.
The contribution of every other state in the theory is exponentially suppressed. Therefore the partition 
function is approximately given by
\begin{align}
 Z(K) &\approx \underbrace{\sum_{(i,\textit{\={\i}})} {\rm Tr}\left(d_{A^*{\bf i}}\,d_{A{\bf i}}\right)^K 
S_{i0}S_{\textit{\={\i}}0}}_{\equiv A(K)}~ e^{\frac{\pi^2\,c}{6\delta K}}\,.
\end{align}
The factor $A(K)$ contains the information about the topological interface. For the entanglement entropy 
we then obtain
\begin{align}\label{eq:Sunderway}
 S &= (1-\partial_K) \log Z(K)\big|_{K\rightarrow1} \nonumber\\
   &\approx (1-\partial_K) \left( \frac{\pi^2 c}{6 \delta K}+\log A(K)\right)\bigg|_{K\rightarrow1}  \\
   &= \frac{c}{3}\,\log L \,+\, \left[\log A(1) - \frac{A'(1)}{A(1)}\right]\,.\nonumber
\end{align}
In the last line we have used~\eqref{def:delta}, and a prime denotes the derivative with respect to $K$.
Note that  time in the channel described in~\eqref{eq:ZKaftermodularS} runs parallel to the interface. 
From \eqref{eq:integralitycondition} we therefore find that 
\begin{equation}
A(1)={\cal N}_{{\bf 0}A}^{\;\;A}
\end{equation} 
is a non-negative integer. It is the multiplicity of the vacuum representation in the twisted torus partition 
function in the channel where time evolves along the interface and its conjugate. If the interface is elementary
we have $A(1)=1$. For the derivative of $A(K)$ one obtains 
$$A'(1) =  \sum_{(i,\textit{\={\i}})} S_{i0}S_{\textit{\={\i}}0}\, {\rm Tr}
\left(d_{A^*{\bf i}}\,d_{A{\bf i}}\right)\, \log \left(d_{A^*{\bf i}}\,d_{A{\bf i}}\right) \,.$$ 
Inserting this in \eqref{eq:Sunderway}, the entanglement entropy becomes
\begin{equation}\label{eq:generalresult}
S = \frac{c}{3}\, \log \, L + \left[\log {\cal N}_{{\bf 0}A}^{\;\;A}  - 
\frac{1}{{\cal N}_{{\bf 0}A}^{\;\;A}}\sum_{(i,\textit{\={\i}})} S_{i0}S_{\textit{\={\i}}0}\, 
{\rm Tr}\left(d_{A^*{\bf i}}\,d_{A{\bf i}}\right)\, \log \left(d_{A^*{\bf i}}\,d_{A{\bf i}}\right) \, \right]\,.
\end{equation}
Within CFT1 we now define
\begin{equation}\label{eq:probability}
p_{(\ii,\alpha\alpha')}^{A}=\frac{d_{A^*{\bf i}}\,d_{A{\bf i}}\,
S_{i0}S_{\textit{\={\i}}0}}{{\cal N}_{{\bf 0}A}^{\;\;A}}\,,
\end{equation}
where the multiplicity labels $\alpha$ and $\alpha'$ both run from $1$ to $M_{\ii}^1$.
For every pair $(i,\textit{\={\i}})$, the matrix $p_{{\bf i}}^{A}\equiv p_{(\ii,\alpha\alpha')}^{A}$
is a positive-semidefinite Hermitian matrix\footnote{Recall that in  unitary theories $S_{i0}>0$}, {\it i.e.} the 
eigenvalues of the $p_{{\bf i}}^{A}$ are
real and positive. Moreover, by \eqref{eq:integralitycondition} we have
\begin{equation}\label{eq:distributionnormaisation}
\sum_{(i,\textit{\={\i}})}{\rm Tr}\, p_{{\bf i}}^A=1\,.
\end{equation}
The set of all eigenvalues therefore forms a probability distribution. In a quantisation where time runs
orthogonally to the interface, the value of ${\rm Tr }\,p_{{\bf i}}^A$ is the probability of 
finding the system CFT1 in the Ishibashi-type state associated to the sector $(i,\textit{\={\i}})$, 
after tracing out CFT2.\footnote{For an interpretation along these lines in terms of a three-dimensional 
topological field theory see \cite{2012PhRvL.108s6402Q}.} Such a state is thermal within its 
sector, and the set of $p_{{\bf i}}^A$ should therefore be understood as defining a reduced density matrix.
Observe that the distribution corresponding to the identity defect in CFT1 is given by
\begin{equation}\label{eq:identitydistribution}
p_{\bf i}^{id} = S_{i0}S_{\textit{\={\i}}0}\,\delta_{\alpha\alpha'}\,,\qquad
\alpha,\alpha'=1,\,2,\ldots,M_{\ii}^1\,.
\end{equation}
Equation \eqref{eq:generalresult} can now be written as
\begin{equation}\label{eq:generalresult2a}
S = \frac{c}{3}\log L\,-\,\sum_{(i,\textit{\={\i}})}{\rm Tr}\, p_{\bf i}^A\,
\log\frac{p_{\bf i}^A}{p_{\bf i}^{id}}\,.
\end{equation}
This is our main result of this section. The quantity
\begin{equation}\label{eq:generalresult2b}
s(\I_A) := -\,\sum_{(i,\textit{\={\i}})}{\rm Tr}\,p_{\bf i}^A\,
\log\frac{p_{\bf i}^A}{p_{\bf i}^{id}}
\end{equation}
is the negative of the relative entropy --- the Kullback-Leibler divergence \cite{KullbackLeibler} --- 
of the probability distribution 
associated to $\I_A$ on the CFT1 side, measured with respect to the probability distribution 
associated to the identity defect $\D_{id}$ of CFT1. One interpretation of this quantity is the 
amount of information lost when the probability distribution is wrongly assumed to be given 
by $\D_{id}$, while it is in reality given by $\I_A$.\footnote{The fact that we find a relative 
entropy ties in nicely with the results \cite{Faulkner:2015csl}.
There it was observed that in higher dimensions, a perturbation of the shape of the entangled 
region leads to a shift in the entanglement entropy given by the Kullback-Leibler divergence
of the probability distributions defined by the reduced density matrices before and after the perturbation.
While the shape does not play a role in our setup, the shift in the entanglement entropy is again 
associated with the difference of the involved density matrices.}

The relative entropy is always non-negative, and vanishes only if the compared probability 
distributions agree.\footnote{Continuous distributions have to agree almost everywhere.} 
Therefore we have $s(\I_A)\leq 0$, which corresponds to the intuition that an interface 
cannot enhance the transmissivity beyond the one of the identity defect in CFT1. 
We have $s(\I_A)=0$ if and only if $p^A=p^{\rm id}$. This 
is the case precisely if $d_{A^*{\bf i}}d_{A{\bf i}}$ is the identity matrix for all pairs 
of representations $(i,\textit{\={\i}})$ which appear in CFT1. A necessary requirement for 
the existence of an interface with this property is that the representation multiplicities 
of CFT2 must not be smaller than those of CFT1. Since both CFTs are unitary and have a single 
vacuum state, modular invariance in fact forces CFT1 and CFT2 to have identical
spectra. Since the necessary condition $d_{A^*{\bf i}}d_{A{\bf i}}=1$ then means that the 
fusion product of the defect and its conjugate is the identity, we have
\begin{equation}
\exists \,\I\,:\,s(\I)=0\quad\Leftrightarrow\quad
Z_{{\rm CFT}1}=Z_{{\rm CFT}2}\,\quad{\rm and}\quad\I^\dagger\I=\D_{id} \quad{\rm in\;CFT}1\,.
\end{equation}
For general CFT1 and CFT2 we may give a simple upper bound for $s$, based on the restricted 
data we have been employing
so far. Without loss of generality every interface $\I$ between 
CFT1 and CFT2 can be associated with a set of diagonal matrices $p_{\bf i}$. Each of these 
matrices $p_{\bf i}$ has at most \mbox{$T_{\ii}={\rm min}(M_{\ii}^1\,,M_{\ii}^2)$} eigenvalues 
different from 0. Varying the remaining eigenvalues we look for the maximal value of $s$ under 
the linear constraint \eqref{eq:distributionnormaisation}. This is only one constraint 
out of the set \eqref{eq:integralitycondition}, such that this calculation will obviously lead
to an upper bound. A maximal value of $s$ would be achieved for the distribution
\begin{equation}\label{eq:limitdistribution}
p_{\bf i}={\rm diag}(p_{(\ii,1)},\ldots, p_{(\ii,T_{\ii})},0,\ldots,0)\quad
{\rm with}\quad 
p_{(\ii,\alpha)}=\frac{S_{i0}S_{\textit{\={\i}}0}}{\sum_{(j,\textit{\={\j}})}T_{\jj}S_{j0}S_{\textit{\={\j}}0}}\,.
\end{equation}
This distribution yields the upper bound
\begin{equation}\label{eq:bound}
s\leq \log\left(\sum_{(i,\textit{\={\i}})}T_{\ii}\,S_{i0}S_{\textit{\={\i}}0}\right)\,.
\end{equation}
The bound is strictly smaller than zero if there is at least one $(i,\textit{\={\i}})$ 
with $T_{\ii}<M_{\ii}^1$. As we have seen above, this is equivalent to having at least one pair 
$(i,\textit{\={\i}})$ where $M_{\ii}^1\neq M_{\ii}^2$. The bound \eqref{eq:bound} is zero if and only if the 
theories CFT1 and CFT2 have the same spectrum. In cases where the CFTs on the two sides are identical, the 
distribution \eqref{eq:limitdistribution} is in particular obtained from the identity defect. 

We emphasise that different interfaces can lead to the same distribution $\eqref{eq:probability}$,
and thus to the same entanglement entropy. In particular, fusing any interface on either side with 
a symmetry defect of the respective theory will leave the distribution unaltered. The reference 
distribution $p^{\rm id}$ of \eqref{eq:identitydistribution} is therefore also obtained from any symmetry
defect in CFT1. On the other hand, every defect whose fusion product with a particular topological
interface leaves the probability distribution of the interface unaltered is a symmetry defect. 

The distributions also do not change if we superpose the same interface multiple times. This is 
obvious from the interpretation of the probability distribution mentioned above. In agreement to
our remark in section~\ref{sec:ReplicaTrickandI}, an interface $\I$ formally has the same probability 
distribution as $M\I$ for any rescaling $M\in\mathbb{C}^*$, and therefore in particular for superpositions
of the same interface. However, the change in the entanglement 
entropy is difficult to compute for general superposition and fusion. This is so because it is in
general difficult to see how closely the probability distribution of the resulting interface 
follows~$p_{\bf i}^{id}$.

For concreteness, let us now consider the defects \eqref{RCFTdefects} in a rational theory with diagonal modular 
invariant. By \eqref{eq:probability}, the interfaces~\eqref{RCFTdefects} of diagonal RCFTs lead to a probability 
distribution
\begin{equation}\label{eq:probabilityrationaldefects}
p_i^a=|S_{ia}|^2\,.
\end{equation}
From our result \eqref{eq:generalresult2a} we therefore obtain the entanglement entropy
\begin{equation}\label{eq:EErationaldefects}
S=\frac{c}{3} \log L - \sum_i  |S_{ia}|^2\,\log\left|\frac{S_{ia}}{S_{i0}}\right|^2\,.
\end{equation}

\subsection*{Example 1: Duality interfaces}

As a class of examples we consider the duality interfaces~\eqref{eq:dualitydefect}.
Here $\I \, \I^\dagger$ projects the theory onto a sector invariant under a symmetry group $G$. 
Invariant states pick up a constant prefactor of $|G|$, the order of the group. 
On the level of equation~\eqref{eq:master2} this means that
\begin{equation}
Z(K) = \text{Tr}\left((\I\I^\dagger)^K e^{-2\delta H K}\right)=  \text{Tr}\left((\oplus_{g\in G} \D_g)^K 
e^{-2\delta H K}\right) =  |G|^K \ \text{Tr}_{inv}\left( e^{-2\delta H K}\right) \ ,
\end{equation}
where in the last line the trace is taken only over the invariant subsector of the initial Hilbert space. 
This partition function is a projection of an initial partition function, which is in line with the fact 
that correlators of invariant fields  in orbifold theories are obtained by projection from the initial theory.
The prefactor $|G|^K$ will drop out in the calculation of the entanglement entropy, so that effectively we 
consider the entanglement of an initial system with a projected system. However, in comparison with the system 
with only the trivial defect inserted, the projection contains a  factor of $|G|^{-1}$ that leads to a shift in 
the entanglement entropy for duality interfaces. The entanglement entropy in the presence of such a duality 
interface is therefore
\begin{equation}\label{eq:dualityEE}
S=\frac{c}{3} \log L - \log |G| \, .
\end{equation}
In terms of the probability distributions introduced earlier we find for the duality defects
\begin{equation}
p_{\bf i}^{duality}= p_{\bf i}^{id} |G| \ {\rm for} \ {\bf i} \ {\rm invariant}, 
\quad  p_{\bf i}^{duality}= 0 \ {\rm otherwise} \ .
\end{equation}
The shift in the entanglement entropy encodes the information loss under a projection.

\subsection*{Example 2: Ising model}

The critical Ising model is described by three primaries $id,\,\epsilon,\,\sigma $. 
It is an example of a diagonal rational theory. The $S$ matrix of the Ising model 
is given by
\begin{equation}
 S_{ij} = \frac{1}{2}\begin{pmatrix}
  1&1&\sqrt{2}\\
  1&1&-\sqrt{2}\\
  \sqrt2&-\sqrt2&0
 \end{pmatrix}\,,~~~\text{with}~i,j\in\{id,\epsilon,\sigma\}\,.
\end{equation}
\noindent
The three elementary topological defects of the Ising model are therefore
\begin{eqnarray*}
\D_{id}&=& \| id\| + \|\epsilon\| + \| \sigma\|\,, \\
\D_{\epsilon}&=& \| id\| + \|\epsilon\|  - \| \sigma\|\,, \\
\D_{\sigma}&=& \sqrt{2}\|id\| - \sqrt{2} \|\epsilon\|  \,.
\end{eqnarray*}
The defect corresponding to the vacuum $id$ is the identity defect. The defect $\D_{\epsilon}$
is a symmetry defect implementing the ${\mathbb Z}_2$ symmetry of the Ising model. The presence of 
these two defects does not result in a shift of the entanglement entropy. The third defect 
$\D_\sigma$ implements  Kramers-Wannier duality. It satisfies the fusion rules
\begin{equation}
\D_\sigma \D_\sigma = \D_{id} + \D_\epsilon\,.
\end{equation}
From our formula \eqref{eq:EErationaldefects} we deduce that the entanglement entropy of $\D_{\sigma}$ is
\begin{equation}
S(\sigma)=\frac{c}{3} \log L - \log 2 \, ,
\end{equation}
which also agrees with the result (\ref{eq:dualityEE}) for duality interfaces where the order of the 
group is~$2$. The result also reproduces the constant shift in the entanglement entropy observed 
in~\cite{Brehm:2015lja}.

\subsection*{Example 3: $su(2)_k$ interfaces and the large $k$ limit}

The diagonal WZW model based on the chiral algebra $su(2)$ at level $k$ has irreducible
representations labelled by half-integer spins $s$. Using the index convention $i=2 s$, 
the integer label $i$ runs from 0 to $k$. The modular $S$ matrix is given by
\begin{equation}
S_{ij}=\sqrt{\frac{2}{k+2}}\sin\left(\frac{\pi(i+1)(j+1)}{k+2}\right)\,.
\end{equation}
By \eqref{eq:EErationaldefects}, the entanglement entropy in the presence of an elementary 
defect~$\D_a$ of the form~\eqref{RCFTdefects} reads
\begin{equation}\label{eq:SWZW}
S(\D_a)=\frac{c}{3} \log L - \frac{2}{k+2}\sum_{i=0}^k\sin^2\left(\tfrac{\pi(a+1)(i+1)}{k+2}\right)\,\log
\tfrac{\sin^2\left(\frac{\pi(a+1)(i+1)}{k+2}\right)}{\sin^2\left(\frac{\pi(i+1)}{k+2}\right)}\,.
\end{equation}
Note that the defect $\D_k$ does not change the entanglement entropy, since this defect simply implements the 
${\mathbb Z}_2$-symmetry acting on the representation labels as $a\to k-a$. 

At large $k$ one obtains the WZW model based on $su(2)$. The central 
charge is $c=3$, and the model can be presented in terms of three bosons on a target 
space $S^3$ with non-vanishing $H$-flux at large radius. The ${\mathbb Z}_2$-symmetry corresponds to the 
reflection symmetry of the three-sphere. At any $k$, the theory contains elementary defects $\D_a$ for non-
negative integers $a\leq k$. To find their geometric interpretation, we recall a few facts on the interpretation 
of symmetry preserving boundary states.  Quite generally, symmetry preserving D-branes on group manifolds wrap 
conjugacy classes~\cite{Alekseev:1998mc,Bachas:2000ik}, which can be automorphism-twisted. In particular, the 
symmetry preserving (Cardy-)states of a WZW model wrap ordinary conjugacy classes of the underlying group $G$. To 
give an interpretation to defects, we first use the folding trick to map  defects to permutation boundary 
conditions for the WZW model based on $G\times G$. Geometrically, these branes wrap twisted conjugacy classes 
where the automorphism is the permutation of the two factors, and the conjugacy class of 
$(g_1, g_2) \in G\times G$ takes the form~\cite{FigueroaO'Farrill:2000ei}
\begin{equation}
C_\omega(g_1, g_2) = \left\{(h_1^{-1} g_1 h_2, h_2^{-1} g_2 h_1)\,|\, h_1 \in G_1, h_2 \in G_2 \right\} .
\end{equation}
The multiplication map $m: G\times G \to G$ maps these conjugacy classes to the conjugacy classes in the 
diagonal~$G$. Indeed, the twisted conjugacy classes of $G\times G$ correspond precisely to the pre-images of the 
conjugacy classes of $G$ under the multiplication map~\cite{FigueroaO'Farrill:2000ei}. In the case of $SU(2)$  
they take the form $S^3 \times S^2$, as the regular untwisted conjugacy classes of $SU(2)$ are generically 
isomorphic to~$S^2$. The conjugacy classes of $\pm 1$ are special and correspond to points. This gives a 
geometric interpretation to the fact that the defects $\D_a$ carry the same labels as Cardy boundary states. 
Indeed, the label $a$ corresponds to a polar angle distinguishing the different 2-spheres $S^2 \subset S^3$ of a
single~$SU(2)$. 

We first compute the entanglement entropy in the large $k$ limit while keeping the label $a$ 
fixed.
In the limit  $k\to \infty$, the correction $s(\D_a)$ to the universal bulk entanglement entropy $\log L$ 
obtained from the defect $\D_a$ becomes an integral,
\begin{align}\label{integration}
s(\D_a)&=-\lim_{k\rightarrow\infty}\frac{2}{k+2}\sum_{i=0}^k\sin^2\left(\tfrac{\pi(a+1)(i+1)}{k+2}\right)\,\log
\tfrac{\sin^2\left(\frac{\pi(a+1)(i+1)}{k+2}\right)}{\sin^2\left(\frac{\pi(i+1)}{k+2}\right)}\nonumber\\
&=-2\int_0^1dx\,\sin^2(\pi(a+1)x)\log(\tfrac{\sin^2(\pi(a+1)x)}{\sin^2(\pi x)})\,.
\end{align}
In particular, we see that in the large $k$ limit, the probability distribution of the interface is a continuous 
sine-square distribution
\begin{equation}\label{sine-squaredistribution}
p_a(x) = 2 \sin^2 \pi  (a+1) x, \quad x \in [0,1]\,.
\end{equation}
The distributions of sphere-like conjugacy 
classes are related to a conjugacy class corresponding to a point.  

The integration can be performed by elementary methods. We first split the logarithmic 
term. The first of the two resulting summands,
\begin{align}\label{step1}
\int_0^1\!\! dx\,\sin^2(\pi(a+1)x)\log(\sin^2(\pi(a+1)x))=
\frac{1}{\pi}\int_0^\pi \!\!dy\,\sin^2y\log(\sin^2y)=\frac{1}{2}-\log2
\end{align}
is independent of $a$. In the other summand we use $2\sin^2x=1-\cos(2x)$
to obtain
\begin{align}\label{step2}
-&\int_0^1dx\,\sin^2(\pi(a+1)x)\log(\sin^2(\pi x))=\nonumber\\
&\qquad\qquad =-\frac{1}{2\pi}\int_0^\pi dy\,\log(\sin^2 y)
\,+\,\frac{1}{2\pi}\int_0^\pi dy\,\cos(2(a+1)y)\log(\sin^2 y)\,.
\end{align}
The first integral on the right-hand side of~\eqref{step2} yields 
$-\int_0^\pi dy\,\log\sin^2 y=2\pi\log2$. In the second integral we use partial
integration to obtain
\begin{equation}\label{step3}
\begin{split}
\int_0^\pi dy\,\cos&(2(a+1)y)\log(\sin^2 y)=\\&=-\frac{1}{a+1}\int_0^\pi dy\,
\sin(2(a+1)y)\cot(y)=-\frac{\pi}{a+1}\,.
\end{split}
\end{equation}
Using \eqref{step1} -- \eqref{step3}, \eqref{integration} becomes
\begin{equation}
s(\D_a)=-\frac{a}{a+1}\,, \qquad a \ll k \,.\label{eq:sDainf}
\end{equation}
In particular, the contribution to the entanglement entropy from such an elementary defect
$\D_a$ is given by a rational number.

However, there is a second class of defects, for which the approximations made in the calculation leading 
to the result \eqref{eq:sDainf} do not hold. This is in particular the case if we pick $a$ such that $a+1$ 
divides $k+2$ and take the limit keeping the ratio $(a+1)/(k+2)$ fixed.
Let us for example consider the case   $a=k/2$ ($k$ even), geometrically 
corresponding to the equatorial two-sphere, which is the fixed point under the involution $a \to k-a$.  
In this case the probabilities $p_i^a$ vanish for $i$ odd, and take the value $2/(k+2)$ for $i$ even. 
Using similar methods as above, the entanglement entropy in the limit $k\to \infty$ becomes
\begin{equation}\label{sDkhalf}
s(\D_{\frac{k}{2}}) = - \log 4\,, \qquad k\to \infty \ .
\end{equation}
Since $\log 4 >1$ one observes that $-s(\D_{\frac{k}{2}})$ deviates substantially from the value \eqref{eq:sDainf}. 
In fact, plotting of $-s(\D_a)$ at finite even $k$ one observes  a peak in the entanglement at $a=k/2$.
Similar, less pronounced peaks are obtained at other values where $a+1$ divides $k+2$. 

For generic defects, $a+1$ does not divide $k+2$, but of course $(a+1)/(k+2)$ is still  a rational number 
that we denote $l/n$, where $l$, $n$ are coprime. It is natural to ask what happens if instead of $a$ 
(as in the computation leading to  \eqref{eq:sDainf}) we keep $l/n$ fixed when taking the large $k$ limit. 
In this case we find from \eqref{eq:SWZW} the 
expression
\begin{equation}\label{sDkrational}
s(\D_{\frac{l(k+2)}{n}-1}) = -\log(2n) - H(n)\,,\qquad k\rightarrow\infty\,,
\end{equation}
where $H(n)$ is the entropy of a probability distribution $p_{m}=\frac{2}{n}\sin^2(\tfrac{\pi m}{n})$ 
for $m=1,\,2,\ldots,n$,
\begin{equation}
H(n)=\sum_{m=1}^{n}\tfrac{2}{n}\sin^2(\tfrac{\pi m}{n})\,\log\big(\tfrac{2}{n}\sin(\tfrac{\pi m}{n})^{2}\big)\,.
\end{equation}
Note that the values of $s$ in~\eqref{sDkrational} are multiply degenerate, as the right-hand side does not 
depend on $l$. 
The entropies \eqref{sDkrational} are bounded from below by 
$s(D_\frac{k}{2})$, showing again that the defect corresponding to the equatorial two-sphere has minimum 
entanglement entropy.  On the other hand, for $n \gg l$ they quickly approach the value $-1$ from below, such that this asymptotic 
expression in fact comes rather close to the approximation~\eqref{eq:sDainf}. 

\begin{figure}[ht]
 \includegraphics[width=.49\textwidth]{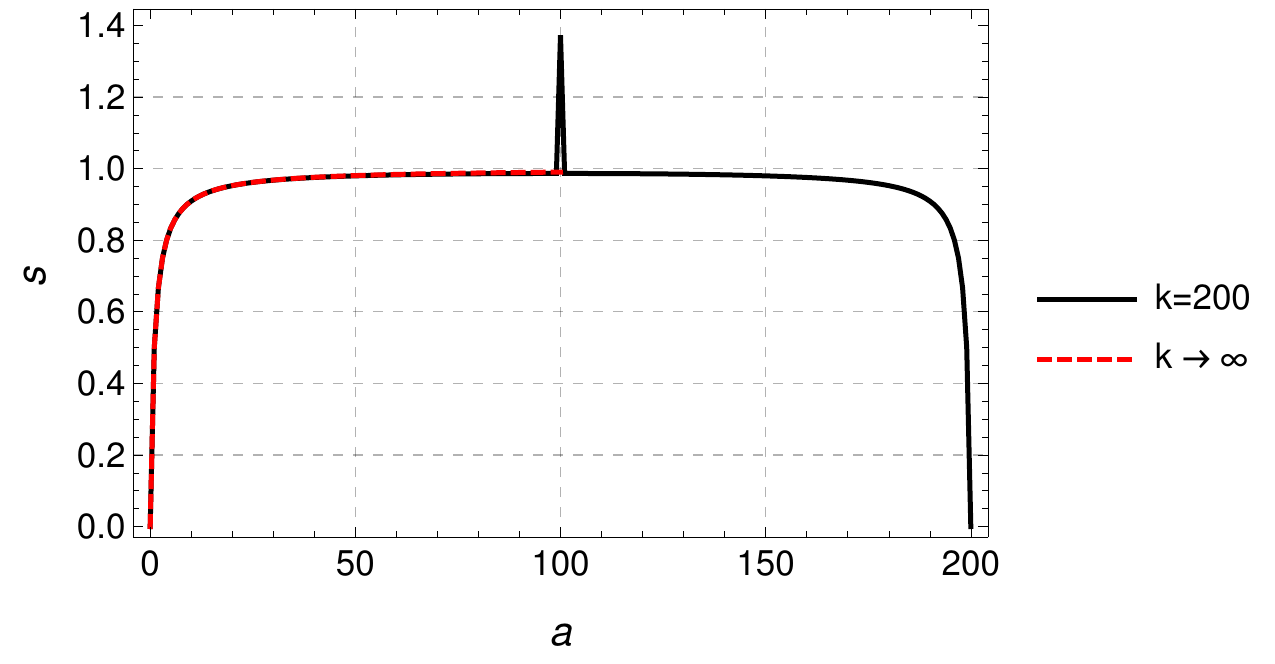}
 \includegraphics[width=.49\textwidth]{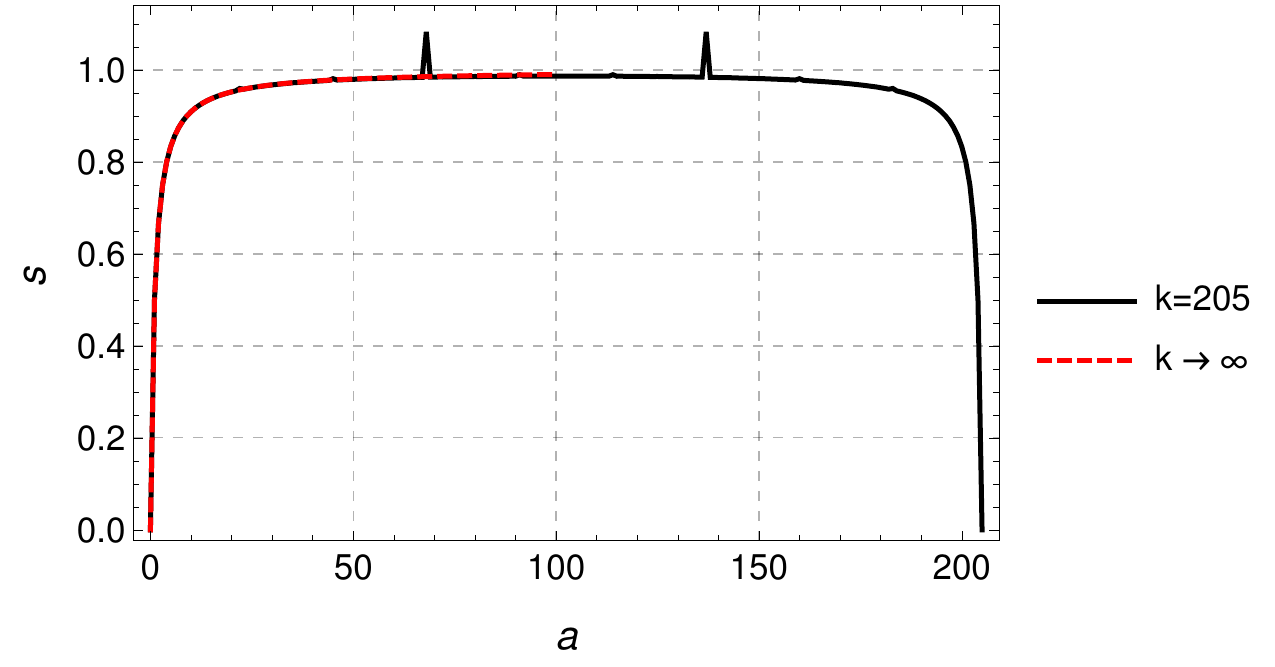}
\caption{Plots of $-s(\D_a)$ for large values of $k$, together with the asymptotic values~\eqref{eq:sDainf}.
The peaks in the plots are captured by the asymptotic expression~\eqref{sDkrational}.}
\label{fig:sDak}
\end{figure}

We will not go much further into details, and instead plot the entanglement entropy correction $-s(\D_a)$ at a 
finite value of $k$ together with the approximation~\eqref{eq:sDainf} in figure~\ref{fig:sDak}. The plot 
illustrates that the values of $s(\D_a)$ approach the asymptotic values~\eqref{eq:sDainf} rather well for 
generic values of $a$. It also illustrates the peaks of the values at
the special points where~\eqref{sDkrational} deviates strongly from~\eqref{eq:sDainf}.

A nice pattern arises when we consider the fusion product of
elementary defects at fixed labels $a$ and $b$ for $k\rightarrow\infty$. For finite $k$, the 
product $\D_{a\times b}=\D_a\D_b$ has the decomposition
\begin{equation}
\D_{a\times b}=\sum_{c}N_{ab}^{\;\;c}\D_{c}
\end{equation}
in terms of elementary defects, where $N_{ab}^{\;\;c}$ are the fusion rules.
Using the fact that in the large $k$ limit the number of vacua on the defects contained in the fusion product is
\begin{equation}
{\cal N}_{0\,a\times b}^{\;\;a\times b}=\min(a,b)+1\,,
\end{equation}
the probability distribution for $\D_{a\times b}$ is given by
\begin{equation}\label{piafusion}
p_i^{a\times b}=\frac{2}{{\rm min}(a,b)+1}\frac{\sin^2(\frac{\pi(a+1)(i+1)}{k+2})
\sin^2(\frac{\pi(b+1)(i+1)}{k+2})}{(k+2)\sin^2(\frac{\pi(i+1)}{k+2})}\,.
\end{equation}
The expression for $s(\D_{a\times b})$ in the large-$k$ limit can be written as
two summands, by splitting off the part involving the logarithm of 
the factor $\min(a,b)+1$ in \eqref{piafusion}. As explained in appendix~\ref{app:1},
using elementary methods one can show that in all cases 
\begin{equation}\label{eq:fusionproduct}
s(\D_{a\times b})=-\frac{p}{q}\,\,+\log(\min(a,b)+1)\,,
\end{equation}
where $p$ and $q$ are natural numbers depending on the labels
$a$ and $b$. Note that the argument in the logarithm is the number of elementary
defects in the decomposition of the fusion product. However, the fact that this logarithm directly reflects the 
number of elementary defects in the decompositon is true only if each of these elementary defects appears with 
multiplicity~1.

\section{Left/Right Entanglement Entropy}\label{sec:LREErational}

In this section we consider a system with a boundary. As mentioned in the introduction, 
the real-space entanglement entropy of a system with boundary receives a correction
by the boundary entropy $s=\log g$, where $g$ is the universal non-integer ground-state 
degeneracy of~\cite{PhysRevLett.67.161}. The entanglement entropy we are interested in
is the left/right entanglement entropy (LREE) considered before in~\cite{PandoZayas:2014wsa} 
for the free boson and in~\cite{Affleck:2009en,Das:2015oha} for generic CFTs. The two subsytems 
consist of the left- and right-moving part of the Hilbert space. 

As mentioned in section~\ref{sec:TopologicalDefects}, a conformal boundary condition is
the maximally reflective solution to the interface conformality 
condition~\eqref{eq:interfaceconformality}. For a CFT in the upper half-plane, the 
components of the bulk energy-momentum tensor satisfy
\begin{equation}
T= \tilde{T} |_{\mathbb R} \, .
\end{equation}
To each boundary condition one can associate a boundary state. If we map the upper half plane 
to the unit disc, the boundary condition becomes such a boundary state in standard radial 
quantisation. The defining property for a conformal boundary state $\ket B$ is the gluing condition
\begin{equation}\label{gluingcondition}
(L_n - \bar{L}_{-n}) |B\rangle =0 \ .
\end{equation}
This means that the boundary state breaks one half of the conformal charges. Such a boundary 
state is coherent, {\it i.e.} it belongs to an extension of the closed string Hilbert space, 
and it is in particular not normalisable. As before, we decompose the Hilbert space as 
in~\eqref{eq:H-decomposition}. If the boundary state preserves the full chiral algebra, the
gluing condition \eqref{gluingcondition} is supplemented by similar conditions for the
additional generators. Together these gluing conditions can only be solved in a sector 
${\cal H}_i \otimes {\cal H}_{\bar{i}}$ where the two representations in the product are
isomorphic. In this way one obtains the Ishibashi states $\ketb{i}$~\cite{Ishibashi:1989}.
For a pair of a boundary state and a dual boundary state imposed on the edges of an 
annulus, the global constraint of consistency under the analog of the modular $S$ transformation is 
referred to as the Cardy constraint~\cite{Cardy1989581}, or as the open-closed string duality. 
As in the case of topological defects it reduces the linear space of solutions of the local gluing 
conditions to the positive cone of a lattice. The remaining consistent boundary states 
\begin{equation}\label{eq:boundarystate}
|B\rangle = \sum_{\bf i} b_{B{\bf i}} \ketb{{\bf i}}
\end{equation}
are linear combinations of Ishibashi states. In this section, bold-faced indices
\begin{equation}\label{boldindicesbdrycase}
{\bf i}=(i,\alpha,\beta)
\end{equation}
only contain one representation label. The sum in~\eqref{eq:boundarystate} and in the rest 
of this section only runs over representations of the bulk space of states
with $i=\textit{\={\i}}$. The multiplicity labels $\alpha$ and $\beta$ distinguish the
different instances where the representation $i$ appears in the holomorphic and antiholomorphic 
part of the space of states, respectively.

In our original setup, correlators and fields in the BCFT will depend on holomorphic and 
anti-holomorphic coordinates restricted to the upper half plane. Using the doubling trick, we 
regard the dependence on anti-holomorphic coordinates $\bar{z}$ on the upper half plane as a 
dependence on holomorphic coordinate $z^*=\bar{z}$ for mirror fields on the lower half plane. 
This means that we consider a chiral construction on the full plane, where the stress tensor 
is continuous everywhere. The boundary condition is then a topological interface in this chiral 
part of a CFT, located on the real line. 

Unfolding the Ishibashi states one obtains interface-like operators $\|{\bf i}\|$ that project onto 
a specific pair of representations ${\bf i}$.
We therefore associate to the boundary state~\eqref{eq:boundarystate} an 
interface operator
\begin{equation}
\I(B) = \sum_{\bf i} b_{B{\bf i}} \, \|{\bf i}\| \, .
\end{equation}
Our computation of the entanglement entropy now proceeds in analogy with the previous sections. 
The characters in the expression
\begin{equation}
 Z(K) = \sum_i {\rm Tr} \left(b_{B^*{\bf i}}b_{B{\bf i}}\right)^K \Tr_{\mathcal{H}_i}\left(e^{-2\delta H}\right) 
= \sum_i {\rm Tr}\left(b_{B^*{\bf i}}b_{B{\bf i}}\right)^K \chi_i\left(e^{-2\delta K}\right) 
\end{equation}
can be written by means of the modular $S$ matrix as
\begin{equation}
 Z(K) = \sum_i {\rm Tr}\left(b_{B^*{\bf i}}b_{B{\bf i}}\right)^K S_{ij}\,\chi_j\left(e^{-\frac{2\pi^2}{\delta K}}\right)
\approx \sum_i {\rm Tr}\left(b_{B^*{\bf i}}b_{B{\bf i}}\right)^K S_{i0}\, e^{\frac{2\pi^2 c}{\delta K 24}} \, .
\end{equation}
For the general boundary state with open string vacuum multiplicity ${\cal N}_{0B}^{\;\;B}$ we 
write the entanglement entropy again in terms of a probability distribution. The distribution is 
defined by the traces of the matrices
\begin{equation}\label{eq:bdrydistribution}
p_{\bf i}^B=\frac{b_{B^*{\bf i}}b_{B{\bf i}}\,S_{i0}}{{\cal N}_{0B}^{\;B}}\,.
\end{equation}
In \eqref{eq:bdrydistribution} we abuse the index notation in the same way as in the previous chapters --- 
while the indices $\bf i$ on the right-hand side contain one multiplicity label
for holomorphic and one for antiholomorphic representations (and we again suppress the summation
over interior labels), the index $\bf i$ on the left-hand side includes two multiplicity labels of the 
same kind. The LREE attributed to a system with boundary condition $B$ 
then reads
\begin{equation}\label{eq:bdryEE}
S=\frac{c}{6}\log L\,-\,\sum_{i}{\rm Tr}\,p_{\bf i}^B\log\frac{p_{\bf i}^B}{S_{i0}}\,.
\end{equation}
Note that the prefactor of the logarithmic term is one half of the prefactor in the case of a full 
theory with a topological interface. This reproduces the area law result mentioned 
in~\cite{calabrese_entanglement_2009}.

A natural question is whether it is again possible to interpret the result in terms of a Kullback-Leibler 
divergence. However, for interfaces there is a generic ``neutral'' interface 
(the identity defect) with respect to which one can compute the relative entropy. This is no longer the 
case for boundaries, as there is no ``neutral'' boundary on the full plane that could serve as a reference 
point. There will generically always be ``information loss'' when left movers are scattered by the boundary 
into right movers. Exceptional cases occur when the boundary condition is  a permutation boundary condition 
obtained by folding an identity or symmetry defect to a boundary condition for a tensor product of identical CFTs.

Technically, one can try to interpret the denominator in the logarithm of~\eqref{eq:bdryEE} as a 
distribution corresponding to the entries $S_{i0}$ times appropriate identity matrices. However, 
the sum over the traces of these matrices is in general not equal to 1, and therefore not a probility distribution. 
In the cases where it is, we indeed obtain the relative entropy with respect to a permutation boundary state, 
where each $b_{\bf i}$ is a permutation matrix. However, in general we conclude that the interpretation as a 
relative entropy fails in the case of the LREE boundary states.

An immediate consequence of loosing the interpretation of the LREE 
as a relative entropy is that the contribution
\begin{equation}\label{LREEs}
s=-\,\sum_{i}{\rm Tr}\,p_{\bf i}^B\log\frac{p_{\bf i}^B}{S_{i0}}
\end{equation}
is not necessarily negative any more. The technique we applied in the case of interfaces
in section~\ref{sec:EErational} now yields the upper bound
\begin{equation}\label{LREEbound}
s\leq \log\bigg(\sum_{i}S_{i0}\bigg)\,.
\end{equation}
This bound does not need to be negative. As an example consider boundary states in diagonal 
rational models. Elementary boundary states which preserve the rational symmetry are labelled 
by irreducible represenations $b$ of the symmetry algebra. The coefficients $b_{bi}$ of the 
elementary boundary states in this case are 
\begin{equation}\label{eq:elementaryboundaryRCFT}
 b_{bi} = \frac{S_{bi}}{\sqrt{S_{i0}}}\,.
\end{equation}
\noindent
From~\eqref{eq:bdrydistribution} and~\eqref{LREEs} we obtain the LREE
\begin{equation}\label{LREECardystates}
S= \frac{c}{6} \log L - \sum_i |S_{bi}|^2 \log \frac{|S_{bi}|^2 }{S_{i0}}\,.
\end{equation}
\noindent
This reproduces the result obtained previously in \cite{Das:2015oha}. It seems plausible
that all symmetry-preserving boundary states in a diagonal model have the 
LREE of the Cardy brane associated to the identity as an upper bound,
\begin{equation}
s\leq -\sum_i S_{0i}^2\log S_{0i}\,.
\end{equation}
The right-hand side is stricter than the bound \eqref{LREEbound}, and it is always positive.

\subsection*{Example: Ising model}

The LREE for boundary states of the Ising model has been discussed
in~\cite{Das:2015oha}. We quote the results here for illustration. The Cardy states in the
Ising model are explicitly given in terms of Ishibashi states by
\begin{align}
\ket{id}&=\tfrac{1}{\sqrt{2}}\big(\,\ketb{id} + \ketb{\epsilon} + 2^{\frac{1}{4}}\ketb{\sigma} \,\big)\,,\nonumber\\
\ket{\epsilon}&=\tfrac{1}{\sqrt{2}}\big(\,\ketb{id} + \ketb{\epsilon} - 2^{\frac{1}{4}}\ketb{\sigma} \,\big)\,,\\
\ket{\sigma}&=\ketb{id} - \ketb{\epsilon} \,.\nonumber
\end{align}
The contributions to the LREE we obtain from 
\eqref{LREEs} are 
\begin{equation}
s=\frac{3\log 2}{4}\;\;{\rm for}\;\;\ket{id},\;\ket{\epsilon}\,,\qquad{\rm and}\qquad
s=0\;\;{\rm for}\;\;\ket{\sigma}\,.
\end{equation}

\subsection*{Example: $su(2)_k$ boundary states and the $k\rightarrow\infty$ limit}

Analogously to the example of $su(2)_k$ defects in section~\ref{sec:EErational} we
consider the LREE of boundary states in the WZW models 
$su(2)_k$ in the limit $k\rightarrow\infty$. For finite $k$, the theory is diagonal 
and rational, and the formulae of~\cite{Das:2015oha} apply (see~\cite{Schnitzer:2015gpa} 
for a discussion of the LREE in WZW models at finite $k$). The Cardy 
states~\eqref{eq:elementaryboundaryRCFT} are again labelled by spins 
$s=b/2$ for $b=0,\,1,\ldots,k$. From~\eqref{LREECardystates}, the universal 
contribution to the LREE by the state $B_b$ is
\begin{align}\label{eq:EEsu2kbranes}
s(b)&=-\sum_{i=0}^k\tfrac{2}{k+2}\sin^2(\pi(b+1)\tfrac{(i+1)}{k+2})\,
\log\tfrac{\sin^2(\pi(b+1)\frac{i+1}{k+2})}{\sin(\pi\frac{i+1}{k+2})}\;\;
+\log\sqrt{\tfrac{2}{k+2}}\,.
\end{align}
Here we have split off a factor depending only on $k$ from the argument of the logarithm,
and used that $\sum_ip^b_i=1$. Observe that the shift term $-\log(k+2)$ in \eqref{eq:EEsu2kbranes}
has the right form to be identified with (the logarithm of) the radius of the target space. 
The target space of the $su(2)_k$ WZW model is the (fuzzy) sphere $S^3$ at radius $R=\sqrt{k}$.
As in the defect case, the sum in \eqref{eq:EEsu2kbranes} becomes an integral
in the $k\rightarrow\infty$ limit. By the same methods as in 
section~\ref{sec:EErational} we obtain
\begin{equation}\label{eq:largek1}
s(b)=-\frac{2b+1}{2b+2}\,+\,\frac{1}{2}\log 2\,+\,\log R\,,\qquad k\rightarrow\infty,\,.
\end{equation}
The positive (and infinite) contribution from the radius is similar
to the radius contribution to the LREE of Dirichlet branes of 
the compactified boson~\cite{PandoZayas:2014wsa}, see~\eqref{PandoZayasResult}.

\subsection*{Example: Fusion of defect and boundary in the $su(2)_k$ WZW model}

To extend the result of the last example we consider the fusion product of an elementary defect operator 
${\cal D}_a$ with a Cardy boundary state $\ket{b}$. This yields a new boundary state 
$\ket{B}={\cal D}_a\ket{b}$. From \eqref{RCFTdefects} and \eqref{eq:elementaryboundaryRCFT} we see 
that the coefficients of $\ket{B}$ are given by
\begin{equation}
b_{Bi}=\frac{S_{ai}^2S_{bi}^2}{S_{0i}^{3/2}}\,.
\end{equation}
The number of open-string vacua in the self-spectrum of $\ket{B}$ is
\begin{equation}
{\cal N}_{0B}^{\;\,B}=\sum |b_{Bi}|^2 S_{0i} =\min(a,b)+1\,,
\end{equation}
as in the case of the fusion products of two elementary defects in example~3 of 
section~\ref{sec:EErational}. The subleading contribution to the LREE can be written as
\be \label{eq:entanglepert} s(a\times b)=\log (\min(a,b)+1)-\frac{\sum_i |b_{Bi}|^2S_{0i} \log|b_{Bi}|^2}{\min(a,b)+1}\,.\ed
Observe that $\min(a,b)+1$ is also again the number of elementary branes in the decomposition
\begin{equation}
\ket B = \sum_{c}N_{ab}^{\;\,c}\ket c
\end{equation}
of the fusion product.

In the large $k$ limit of the $su(2)_k$ WZW model, the LREE of 
the fusion product differs from the entropy of the original boundary state $\ket b$
again by a rational term and the logarithm of the number of elementary branes in the
decomposition.
Indeed, in the limit of large $k$ the numerator in the second term of the right-hand side 
of~\eqref{eq:entanglepert} becomes
\begin{align} \label{eq:integralRa}
&\sum_i |b_{Bi}|^2S_{0i} \log|b_{Bi}|^2\xrightarrow{k\rightarrow\infty}\\
&\quad -\log\sqrt{\tfrac{2}{k+2}}\,+\,\frac{2}{\pi}\int_0^\pi \tfrac{\sin^2((a+1)x)\sin^2((b+1)x)}{\sin^2(x)} 
\log \left( \tfrac{\sin^2((a+1)x)\sin^2((b+1)x)}{\sin^3(x)}\right)\ dx\,. \nonumber
\end{align}
A similar calculation as in example~3 of section~\ref{sec:EErational} leads to
\be \label{eq:spertlargek} s(a\times b)=\log \mbox{min}(a+1,b+1) +\frac{1}{2}\log 2+\log R - \frac{p}{q} \,,
\qquad k\rightarrow\infty\,,\ed
for some $p,\,q\in\mathbb{N}$.
The difference between the entanglement entropy of the boundary state after fusion \eqref{eq:spertlargek} 
and the original boundary state \eqref{eq:largek1} for $k=\infty$ is therefore
\be \label{eq:deltas} s(a\times b)-s(b)=\log (\mbox{min}(a,b)+1)-\frac{p}{q}\,.
\ed

\section{Results for Bosonic Tori}\label{sec:EEtori}
\subsection{EE through topological defects}

For the case of $d$ free bosons compactified on a torus the interface operators are explicitly 
known~\cite{bachas_worldsheet_2012}. The ground states of the theory form an even, self dual 
lattice $\Gamma \subset \R^{d,d}$. The lattice vectors are of the form $\gamma=(p, \bar{p})$, 
where the $d$-dimensional vectors $p$ and $\bar p$ denote left- and right-moving momenta. 
We will consider topological interfaces that also preserve the full  $u(1)^d$ symmetry. 
These interfaces are specified by a gluing matrix $\Lambda \in O(d|\R)\times O(d|\R)$.  
Similarly to the rational case discussed earlier, the interface operators can be written 
as linear combinations of operators between $u(1)^d$ highest weight representations:
\begin{equation}
\I_{12}(\Lambda) = \sum_{\gamma \in \Gamma^\Lambda_{12}} d_{\Lambda \gamma} || \gamma || \,.
\end{equation}
As before, $||\gamma ||$ is an intertwiner of the
representation space specified by the lattice vector $\gamma$, and $d_{\Lambda \gamma}$ are 
prefactors constrained by consistency under modular $S$ transformation. The range of the summation is 
restricted to a sublattice, given in terms of a gluing condition $\Lambda$ for the lattices $\Gamma_1$ 
and $\Gamma_2$ on the two sides of the interface,
\begin{equation}
\Gamma_{12}^\Lambda = \{\gamma \in \Gamma_1 \,|\,\Lambda\gamma\in \Gamma_2\} =
\Gamma_1 \cap \Lambda^{-1} \Gamma_2  \subset \Gamma_1.
\end{equation} 
For admissible gluing conditions $\Lambda$, the sublattice $\Gamma_{12}^\Lambda$ 
has full rank.
Consistency under modular $S$ transformation then demands that 
$d_{\Lambda \gamma} = g_{12}^\Lambda \exp (2\pi i \varphi (\gamma) )$, where  
$\varphi\in (\Gamma^\Lambda_{12})^\star$ and 
\be(g_{12}^\Lambda)^2 = |\Gamma_1/\Gamma_{12}^\Lambda|\,\label{eq:gI}\ed  
is the index of the sublattice $\Gamma_{12}^\Lambda$ inside the lattice $\Gamma_1$.
The topological interface operator splits into a lattice and an oscillator part,
\begin{equation}
 \I_{12} = \I^0_{12}(\Lambda) \prod_{n>0} \I^n_{12}(\Lambda)\,,
\end{equation}
where
\begin{equation}
\I^0_{12} = g_{12}^\Lambda \sum_{ \gamma \in \Gamma_{12}^\Lambda} 
e^{2\pi i\varphi(\gamma)} \ket{\Lambda\gamma}\bra{\gamma}
\end{equation}
gives the map for the zero modes and the
\begin{equation}
\I^n_{12} = \exp\left(-\frac{1}{n}\left( a^2_{-n}\Lambda_{11}a_n^1 + 
\tilde a_{-n}^2 \Lambda_{22}\tilde a_n^1\right) \right)
\end{equation}
for $n>0$ give the contribution of the higher modes. It is implicitly understood that 
modes of CFT1 act from the right and modes of CFT2 from the left of $\I^0_{12}$. 

In order to determine the entanglement entropy we proceed as before.
The partition function of the $K$-sheeted Riemann surface for the topological 
defect~\eqref{eq:master2} is
\begin{equation}
\begin{split}
 Z(K) &= \Tr \left( \left(\I \I^\dagger\right)^K e^{-2\delta K H} \right)\\ 
&= (g_{12}^\Lambda)^{2K} \sum_{(p,\bar{p})=\gamma\in\Gamma^\Lambda_{12}}  
\chi_p(i\frac{\delta}{\pi} K )\bar \chi_{\bar{p}}(i\frac{\delta}{\pi} K )\,,
\end{split}
\end{equation}
where the $\chi_p$ are the $u(1)$ characters. 
We perform a modular $S$ transformation and express $Z(K)$ in terms of characters depending on the 
variable~$i\pi/K\delta$. This leads to a summation over 
lattice vectors in the dual lattice $\Gamma_{12}^\vee$. In the limit 
$\delta \ll 1$ we approximate the lattice sum by the dominant contribution of the vacuum
$p=\bar{p}=0$,
\begin{equation}
\begin{split}
Z(K) &=  (g_{12}^\Lambda)^{2K} \sum_{(q,\bar{q})\in\Gamma_{12}^\vee} a_{(q,\bar{q})} 
\chi_q(i\frac{\pi}{K\delta}  )\bar \chi_{\bar{q}}(-i\frac{\pi}{K\delta}  ) \\
&\approx  (g_{12}^\Lambda)^{2K} a_{(0,0)}  e^{\frac{\pi^2\,d}{6\delta K}}\,.
\end{split}
\end{equation}
We now use that the interfaces with the normalization \eqref{eq:gI} are elementary. 
For $K=1$, we obtain the ordinary defect partition function, where the multiplicity 
of the vacuum propagating in the dual channel is $1$. 
We therefore conclude that $a_{(0,0)}=1/ (g_{12}^\Lambda)^{2}$.

Using $\delta = \pi^2/\log(L)$, $c=d$ for the central charge of $d$ bosons, 
and~\eqref{eq:gI}, the entanglement entropy is given by
\begin{equation}\label{eq:EEtori}
 S = (1-\partial_K)\log(Z(K))\big|_{K=1} = \frac{c}{3}\log(L) - 
\log|\Gamma_1/\Gamma_{12}^\Lambda|\,.
\end{equation}
In the special case $d=1$, {\it i.e.} for a free boson compactified on a circle, 
conformal interfaces between CFT1 and CFT2 are classified by two winding numbers 
$k_1$ and $k_2$. For generic compactification radii these interfaces are not topological, 
but they become so by choosing radii to satisfy the relation $R_1/R_2 = k_2/k_1$~\cite{Bachas:2007td}.  
In this case the index of the sublattice is~\cite{Bachas:2007td, bachas_worldsheet_2012} 
\begin{equation}
|\Gamma_1/\Gamma_{12}^\Lambda| = |k_1 k_2|\, ,
\end{equation}
such that the entanglement entropy through the topological interface is given by 
\be S=\frac{1}{3}\log(L)-\log |k_1k_2| \ ,\ed
in agreement with the result of~\cite{sakai_entanglement_2008}. 

All topological toroidal interfaces are duality interfaces according to our previous 
definitions. A subclass of them are symmetry interfaces and describe automorphisms 
of the toroidal CFT. They are associated to gluing matrices in the T-duality 
group~$ O(d,d,\Z)$. In particular, for those matrices we get 
$\Gamma_{12}^\Lambda = \Gamma_1$, which means that the defect couples to the full
momentum lattice and no ground states are projected out. In this case, there is no 
contribution to the entanglement entropy from the interface.

The broader class of duality interfaces is specified by matrices in 
$O(d,d,\mathbb{Q})$. These interfaces can in particular be related to orbifold 
constructions. In the case of  a single circle, where 
$R_1/R_2 = k_2/k_1$~\cite{Bachas:2007td,bachas_worldsheet_2012}, the theory with 
radius $R_1$ can be obtained from the theory with radius $R_2$ by orbifolding with respect
to the shift symmetry
\be X \mapsto X+2\pi R_1\,. \ed
The orbifold group generated by this symmetry is $\Z_{|k_1 k_2|}$, and hence of order $|k_1 k_2|$. 
It is clear that $\I\I^\dagger$ projects the theory with $R_2$ onto the sector invariant 
under the orbifold group. We see that for circle theories the contribution to the 
sub-leading term of the entanglement entropy is set by the order of this orbifold group,
\be S=\frac{1}{3}\log(L)-\log |G| \,,\ed
in agreement with the general result \eqref{eq:dualityEE}.

\subsection{LREE of bosonic tori}

We consider the LREE for $d$ free bosons compactified on a torus. 
The gluing conditions can be written as~\cite{bachas_worldsheet_2012,blumenhagen_basic_2013}
\be\label{gluconlree}(a_n+O\tilde{a}_{-n})\ket{B}=0\,,\ed
where $O\in O(d|\R)$. The ground states solving the zero mode condition are given by
\be \label{gcLREEtorus}\Gamma^O=
\left\{ \begin{pmatrix}
-Ox\\
x
\end{pmatrix}\cap \Gamma\,~|~ x \in \R^d \right\}\,,\ed
where $\Gamma$ is the charge lattice of the torus model. The boundary state is a superposition of 
Ishibashi states $\ketb{p,\bar{p}}$ built on ground states $(p, \bar{p}) \in \Gamma^O$,
\be \ket{B}={g}\sum_{(p,\bar{p})\in\Gamma^O}e^{i\varphi(p,\bar{p})}\ketb{p,\bar{p}}\,,\ed
where the function $\varphi \in (\Gamma^O)^*$ specifies  the $D$-brane moduli and the 
$g$-factor is fixed by the condition that the open-string vacuum appears with multiplicity~1. 
As before, we unfold the boundary state and associate an 
interface  between left- and right-movers of the free boson theory. For this, we introduce the 
projections $\pi (\Gamma^O)$ and $\bar\pi (\Gamma^O)$ of the lattice $\Gamma^O$ to the left- 
and right-moving parts respectively.
On the level of ground states, the interfaces therefore maps 
$\pi (\Gamma^O) \ni p \to -Op \in \bar\pi (\Gamma^O)$. 
The $g$-factor is given by
\begin{equation}\label{gfactorvolume}
g= \mathit{vol}(\pi(\Gamma^O)) ,
\end{equation}
the volume of the unit cell of $\pi(\Gamma^O)$. 
\noindent
The computation of the entanglement entropy now proceeds  in analogy to the case of 
topological interfaces. The partition functions on the $K-$sheeted torus are approximated by
\begin{equation}
 Z(K) = g^{2K} \sum_{(p,\bar{p})\in\Gamma^O}  \,\chi_p(i\frac{\delta }{\pi} K) \,
\xrightarrow{\;S \text{ trsf}, \,\delta\ll1\;}\, g^{2K-2} e^{\frac{\pi^2}{12\delta K}}\, ,
\end{equation}
where we used again that the vacuum in the open string channel has multiplicity~1 for $K=1$.
For $c=d, \delta = \pi^2/\log(L)$ we obtain from this the LREE
\begin{equation}\label{eq:LREEtori}
 S = \frac{c}{6} \log(L) - \log \mathit{vol}(\pi(\Gamma^O)) . 
\end{equation}
The subleading part of the LREE is determined by the $g$ factor of the boundary state.
To relate this quantity to the torus geometry, let us recall that $\Gamma$ is a Narain lattice
given by
\begin{equation} \label{eq:Narain}
\Gamma= \left\{ \left(\begin{array}{c} \frac{1}{2} E^{-1} N + E^T (1+B)M 
\\ -\frac{1}{2} E^{-1} N + E^T (1-B) M \end{array} \right) \, | \, M,N \in {\mathbb Z}^d  \right\}  \ ,
\end{equation}
where $G=EE^T$ is the metric and $B$ the Kalb-Ramond field on the target space, and $N$ and $M$ are the momentum 
and winding quantum numbers, respectively.
Let us consider a $D1$ brane in $d=2$ dimensions for the geometric case where $B$ is zero. 
If the D1 brane were located in infinite flat space, we would specify the direction of the brane by specifying the 
momenta perpendicular to it; a localisation of the brane to its world-volume direction would then be achieved 
by integration over these momenta. On a torus, the momenta are part of a lattice. 
We can fix our brane by choosing the elementary generator of transverse momenta that couple to the brane to be given by
\begin{equation}\label{eq:p}
N^0= \left( \begin{array}{c} N_1^0 \\ N_2^0 \end{array} \right)\,,
\end{equation}
with two  integers $N_i^0$ that we assume to be relatively prime. This choice determines the winding modes our D1 brane 
can couple to. The elementary winding generator $M^0=(M_1^0,\,M_2^0)$
is again specified by two coprime integers $M_1^0,\,M_2^0$, and have to satisfy the orthogonality constraint
\begin{equation}
N_1^0 M_1^0 + N_2^0 M_2^0=0\,.
\end{equation}
The equation is solved by $M_1^0 = - N_2^0, \, M_2^0 = N_1^0$. By this we have fixed
a D1 brane for which the lattice $\Gamma^O$ is precisely spanned by the two generators 
$N^0$ and $M^0$ for $N$ and $M$ in \eqref{eq:Narain}, respectively. It can be checked explicitly that 
these lattice vectors solve \eqref{gluconlree} with 
\be \label{glumatrix} O=\begin{pmatrix}
\cos(2\theta) & \sin(2\theta)\\
\sin(2\theta) & -\cos(2\theta)
\end{pmatrix}\in O(2,\mathbb{R})\,,\ed
where $\theta=\arctan\big(-(E^{-1}N^0)_1/(E^{-1}N^0)_2\big)$~\cite{Bachas:2007td}.
To compute the $g$-factor we now have to compute the volume of (the unit cell of) this lattice, 
projected to the left-movers. After some algebra one obtains 
\begin{equation}
g^2 = \frac{1}{2 \det E} \left( (N_2^0)^2 G_{11} + (N_1^0)^2 G_{22} -2 G_{12} N_2^0 N_1^0 \right) \ .
\end{equation}
Mapping $N_i^0$ to the winding numbers of the brane $N_1^0 = k_2, N_2^0=-k_1$, we see that
\begin{equation}
g^2 =\frac{\mathit{length}^2}{2 \mathit{vol}} , 
\end{equation}
where $\mathit{length}$ refers to the length of the brane and $ \mathit{vol}$ to the volume of the torus.
This is in agreement with geometrical expectations. Note also, that in the special case of a rectangular 
torus with diagonal metric where the radii are related by a rational number the above result agrees 
with the one for topological interfaces of the previous section. 

The left- right-entanglement entropy has been computed before for the case of a free boson 
in~\cite{PandoZayas:2014wsa}. To compare the results, note that in one dimension the 
left-moving momenta are given by $a_0 = N/2R + MR$ and the right-moving momenta by 
$\bar{a}_0 = N/2R - MR$. The matrix $O$ in the gluing condition reduces to a choice of sign. 
For Dirichlet branes we have $O=1$, and only ground states without winding contribute to the boundary state. 
We therefore have $\Gamma^O = \left\{ (N/2R, N/2R) \right\}$, and the volume of the projected unit cell is~$1/2R$. 
Similar considerations also hold for Neumann branes.
Our result \eqref{eq:LREEtori} for the LREE of a single boson compactified on a circle thus gives
\begin{equation}\label{PandoZayasResult}
 S = \frac{1}{6}\log L - \left\{\begin{matrix}
                                  \log R & \text{for }  O = -1 \text{ (Neumann b.c.)} \\
                                  \log\frac{1}{2R} & \text{for }  O = 1 \text{ (Dirichlet b.c.)}
                                 \end{matrix}\right.\,.
\end{equation}
which in particular reproduces the results of \cite{PandoZayas:2014wsa}.

\section{Conclusion and Outlook}\label{sec:conclusion}

In this paper we have discussed the entanglement entropy through topological interfaces, 
and the left/right entanglement entropy at conformal boundaries. Our focus has been on 
unitary CFTs with discrete spectrum.
For an interface between CFT1 and CFT2 we trace out the half-plane on the CFT2 side of the interface.
The eigenvalues of the resulting reduced density matrix \eqref{eq:probability} give the probability of 
finding the reduced system CFT1 in a thermal state in the representation $(i,\textit{\={\i}}\,)$.   
The general result for the entanglement entropy through a topological interface is obtained 
in~\eqref{eq:generalresult2a}. The universal bulk term proportional to $\log L$ is not affected, 
as correlation functions do not depend on the position or shape of the interface insertion. 
To subleading order the topological interface contributes a universal term to the entanglement entropy.
The contribution is the negative of the relative entropy (Kullback-Leibler divergence) of the 
distribution associated to the interface, compared with the situation when there was no 
interface to start with.

By unfolding a theory with a boundary we obtain a topological interface in a chiral theory,
allowing a derivation of the left/right entanglement entropy \eqref{eq:bdryEE} in analogy 
to the interface case. In this situation we lose the interpretation of the entanglement
entropy in terms of a relative entropy. However, the derivation of the entanglement entropy 
proceeds in analogy to the interface case, and the resulting formulas have a similar structure.

The Kullback-Leibler divergence has many appealing features. In this paper, we have interpreted only some 
of them in the context of defects and interfaces. In particular, we have given a meaning to 
its positivity -- the interface will never increase the entanglement.  We have also connected 
the vanishing of the Kullback-Leibler divergence to properties of interfaces that do not lead to information 
loss. 

One particular issue that we did not touch in this paper is that the Kullback-Leibler divergence measures the difference 
between probability distributions, and in this sense shares some properties with a distance (for an application to 
distances between quantum field theories see {\it e.g.} \cite{Balasubramanian:2014bfa}). In the present paper
we only obtained Kullback-Leibler divergences of interfaces with respect to the identity defect. In order to explore the
distance property, it would be interesting to study a concrete physical realisation 
for Kullback-Leibler divergences with respect to arbitrary interfaces as reference points.

The Kullback-Leibler divergence is evidently not symmetric, and even after a suitable symmetrisation fails to satisfy 
the triangle inequality. The distance measure it might yield for topological defects would therefore share 
the same features.

We recall that interfaces have been used to define distances before. In particular, in \cite{Bachas:2013nxa} 
the interface entropy $\log g$ of deformation interfaces\footnote{Deformation interfaces relate different CFTs in 
the same moduli space in a minimal way.} was identified with Calabi's Diastasis function. Ultimately, the 
proposal was that the $g$-factor between interfaces can be used to define a distance between different 
CFTs. However, also in this case it was observed that the triangle inequality will not be satisfied. 

A further observation for the $\log g$ distance is that it gives rise to a metric at the infinitesimal level, 
which coincides with the Zamolodchikov metric on the moduli space. This is again similar in the 
case of the Kullback-Leibler divergence, where the infinitesimal limit yields the Fisher information metric. At the moment, 
for our purposes this property is at the moment rather formal, as the interfaces we have studied here
are generally labelled by discrete numbers.

Note that the deformation interfaces considered in \cite{Bachas:2013nxa} are in particular not topological.  
In fact, all of them are in the deformation class of the identity defect. The entanglement entropy 
through such interfaces has been investigated only in examples. One particular case is the theory of 
a compactified free boson~\cite{sakai_entanglement_2008}. There the subleading term 
of the entanglement entropy vanishes, and instead the prefactor of the $\log$ term changes. In 
fact, it is suggestive that the subleading terms we discussed in this paper vanish more generally 
for such deformations of the identity. It would be interesting to prove a general statement about 
the form of the entanglement entropy along these lines.

Finally, let us comment on $N=2$ supersymmetric theories. Such theories can be topologically twisted, and
one can formulate boundary as well as interface gluing conditions that are compatible with the topological 
twist. On the level of the topological theory, all interfaces can be moved freely. It would be 
interesting to consider entanglement entropy through topological interfaces in the supersymmetric 
situation, where entanglement should have a topological interpretation. For results on the supersymmetric 
case without interfaces see \cite{Giveon:2015cgs}.

\section*{Acknowledgements}
The authors would like to thank Charles Melby-Thompson and Erik Plauschinn for useful discussions. 
This work was partially supported by the DFG Transregio Grant TRR33 and the Excellence Cluster Universe.

\appendix

\section{Entanglement entropy of a fusion product}\label{app:1}

In this appendix we show that the pattern \eqref{eq:fusionproduct} holds. This follows 
from the fact that in the large~$k$ limit, the contribution
\be \label{eq:longintegral} -2\int_0^1 \frac{\sin^2(\pi(a+1)x)\sin^2(\pi(b+1)x)}{\mbox{min}(a+1,b+1)
\cdot \sin^2(\pi x)}\log \left(\frac{\sin^2(\pi(a+1)x)\sin^2(\pi(b+1)x)}{\sin^4(\pi x)}\right)\ dx\ed
is a rational number. In order to see this, the basic integral that has to be evaluated is
\be \label{eq:fundamental1} -\frac{1}{\pi}\int_0^\pi \frac{\sin^2((a+1)x)\sin^2((b+1)x)}{\sin^2(x)}
\log \left(\sin^2((l+1)x)\right) \ dx\,.\ed
By the two representations of Clausen's function $Cl_1$ we can express the logarithmic factor
in terms of the sum
\begin{equation}\label{ClausenIdentity}
\log(\sin^2(y)) = -\log 4\,-\,\sum_{k=1}^\infty\frac{2\cos(2ky)}{k}\,.
\end{equation}
We note that the $\log 2$ terms cancel out in (\ref{eq:longintegral}), so it is enough to keep
only the sum over cosines from the right-hand side of \eqref{ClausenIdentity}
for further calculations. One might be concerned that while the individual terms in the summation over $k$
give rational results, resummation may yet yield something non-rational. In order to see that this is not the 
case we eliminate the sine functions in the denominator of~\eqref{eq:fundamental1} by writing the remaining sine 
functions in the numerator in terms of spread polynomials \cite{S.Goh} 
\be \label{eq:fundamental3} \sin^2(nx)= \sum_{p=0}^{n-1}\tfrac{n}{n-p}\tbinom{2n-1-p}{p}%
(-4)^{n-1-p}\sin^{2(n-p)}(x)=:\sum_{\{p\}} \sin^{2p}(x)\,.\ed
Since we are not interested in the precise value of the finite result, the only relevant
property for us is that the spread polynomials have rational coefficients and finite order. 
We define the summation symbol on the right-hand side to indicate a finite sum of trigonometric 
functions with rational coefficients. We reduce the right-hand side further by the identity
\begin{equation}\label{eq:fundamental4} 
\sin^{2p}(x)=\tfrac{1}{2^{2p}}\tbinom{2p}{p}
+\tfrac{2}{2^{2p}}\sum_{r=0}^{p-1}(-1)^{p-r}
\tbinom{2p}{r}
\cos(2(n-r)x)=:\sum_{\{r\}} \cos(2rx)\,.
\end{equation}
The purpose of writing \eqref{eq:fundamental3} and \eqref{eq:fundamental4} is to 
demonstrate that~\eqref{eq:longintegral} can indeed be written in the form
\begin{equation}\label{eq:fundamental5} 
\frac{1}{\pi}\int_0^\pi \sum_{\{s\}}\cos(2sx)\, \sum_{k=1}^\infty \cos(2k(l+1)x)\frac{1}{k}\ dx\,.
\end{equation} 
By the integral identity $ \int_0^\pi \cos(nx)\cos(mx)\ dx =\frac{\pi}{2}\delta_{n,m}$,
the finite sum over $s$ reduces the infinite sum over $k$ to a rational result,
and we obtain \eqref{eq:fusionproduct} as proposed.

\bibliographystyle{JHEP}
\bibliography{lit}

\end{document}